\documentclass[a4paper,11pt]{article}
\pdfoutput=1
\usepackage{jcappub}
\usepackage{comment}
\usepackage[T1]{fontenc} 
\usepackage{upgreek}
\usepackage{xspace}
\usepackage{graphicx}
\usepackage[capitalize]{cleveref}
\usepackage{enumitem}
\usepackage{subcaption}
\usepackage[modulo]{lineno}
\usepackage{wrapfig}
\usepackage{indentfirst}
\usepackage[dvipsnames]{xcolor} 
\usepackage[modulo]{lineno}
\usepackage{multirow}
\usepackage[rightcaption]{sidecap}
\usepackage{soul}
\usepackage{tabularx}
\usepackage{booktabs}
\newcolumntype{Y}{>{\centering\arraybackslash}X}
\usepackage[style=nature]{biblatex}
\usepackage[separate-uncertainty]{siunitx}[=v2]

\title{Study of the muon component in the core-corona model using CONEX~3D}

\author[a]{Ana Martina Botti}
\author[]{\hspace{-.2cm},}
\author[b, *]{Isabel Astrid Goos\note[*]{Corresponding author}}
\author[]{\hspace{-.2cm},}
\author[c]{Matias Perlin}
\author[]{\hspace{-.2cm},}
\author[c]{and Tanguy Pierog}

\affiliation[a]{Fermi National Accelerator Laboratory (Fermilab) \\ PO Box 500, Batavia IL, 60510, USA}
\affiliation[b]{Laboratoire Astroparticule et Cosmologie (APC) \\
10 Rue Alice Domon et Léonie Duquet, 75013 Paris, France}
\affiliation[c]{Karlsruher Institut für Technologie (KIT), Institute for Astroparticle Physics, \\
Hermann-von-Helmholtz-Platz 1, 76344 Eggenstein-Leopoldshafen, Germany}

\emailAdd{goos@apc.in2p3.fr}

\abstract{The discrepancy between models and data regarding the muon content in air showers generated by ultra-high energy cosmic rays still needs to be solved. The CONEX simulation framework provides a flexible tool to assess the impact of different interaction properties and thus address the muon puzzle. In this work, we present the multidimensional extension of CONEX and show its performance compared to CORSIKA by discussing muon-related air-shower features for three experiments: KASCADE, IceTop, and the Pierre Auger Observatory. We also implement an effective version of the core-corona model to demonstrate the impact of the core effect, as observed at the LHC, on the muon content in air showers produced by ultra-high energy cosmic rays. At a primary energy of $E_0 = \SI{e19}{\eV}$, we obtain an increase of $15\%$ to $20\%$ in the muon content.}


\begin{document}

\maketitle
\flushbottom

\newcommand{\lna}{\avg{\ln\!A}}
\newcommand{\xmax}{X_\text{max}}
\newcommand{\nch}{N_{\rm ch}}
\newcommand{\nmu}{N_\mu}
\newcommand{\lnn}{\ln\!\nmu}
\newcommand{\fcore}{\omega_\text{core}}
\newcommand{\Esub}[1]{E_{#1}}
\newcommand{\EsubText}[1]{\Esub{\mathrm{{#1}}}}
\newcommand{\QGSJET}{\textsc{QGSJet-01}}
\newcommand{\QGSJETIIc}{\textsc{QGSJetII.03}}
\newcommand{\QGSJETIId}{\textsc{QGSJetII.04}}
\newcommand{\EPOSLHC}{\textsc{EPOS~LHC}}
\newcommand{\SIBYLLd}{\textsc{SIBYLL~2.3d}}
\newcommand{\CONEX}{\textsc{conex}}

\section{The muon puzzle}\label{sec:experiments}

Several riddles remain open in the fields of high and ultra-high energy cosmic rays. For energies above \SI{e15}{\eV}, we can detect cosmic rays only indirectly through the secondary particles, or extensive air showers, produced when they interact with nuclei in the Earth's atmosphere~\cite{engel2011extensive}. The Heitler-Matthews model~\cite{matthews2005heitler, heitler1984quantum} describes particle production in air showers in a simplified way: neutral pions, created along the shower, decay to photons that feed the electromagnetic component (encompassing photons, electrons, and positrons) by subsequent pair creation and Bremsstrahlung processes. Alongside neutral pions, charged pions typically decay into muons, forming the muonic component. 

This model reveals two mass- and energy-sensitive observables~\cite{kampert2012measurements}: the atmospheric depth at the maximum development of the cascade, $X_\mathrm{max}$, mainly determined by the electromagnetic component of the shower, and the number of muons at ground level, $N_\mu$, driven by the shower's hadronic core. Then, we use simulations to interpret air shower measurements and to deduce, for example, the mean logarithmic mass of cosmic rays $\langle \ln{A} \rangle$ from $X_\mathrm{max}$ and $N_\mu$. Most of the uncertainty in $\langle \ln{A} \rangle$ stems from the discrepancies in these models. This uncertainty ultimately challenges our understanding of the astrophysical scenarios in which cosmic rays are produced and accelerated.

Currently, the discrepancy between the muon density measured with surface arrays and that predicted by hadronic interaction models~\cite{albrecht2022muon} represents one of the most pressing questions in air shower physics. Refs.~\cite{albrecht2022muon, Cazon:2020zhx} report on the so-called \textit{muon puzzle}, observed in a broad energy range from about \SI{e16}{\eV} up to \SI{e20}{\eV}, and in experiments measuring showers under different atmospheric conditions with diverse detection techniques. In addition, the number of muons is also sensitive to hadronic multiparticle production at lower energies~\cite{meurer2006muon}, further expanding the energy range that needs to be revised in the models.

Air-shower experiments located at different altitudes measure the shower development at different stages, challenging data interpretation. Following~\cite{albrecht2022muon, Cazon:2020zhx}, we categorized cosmic ray experiments into three groups: the IceCube Neutrino Observatory with its surface array IceTop at around \SI{2800}{\m} a.s.l., the Pierre Auger Observatory and the HiRes/MIA Experiment at ca. \SI{1500}{\m} a.s.l., all other experiments are located between 100 and \SI{250}{\m} a.s.l.. To compare simulations associated with different conditions, we discuss the cases of IceTop at IceCube, the Pierre Auger Observatory, and the KASCADE Experiment~\cite{aab2016pierre, abbasi2013icetop, apel2010kascade}. These cover different altitudes, energy ranges, distances to the shower core, and experimental configurations.

The Pierre Auger Observatory is a hybrid detector comprising a \SI{3000}{\kilo\m\squared} surface array of over 1600 water Cherenkov detectors and 24 fluorescence telescopes overlooking the layout from the periphery~\cite{pierre2020pierre, aab2016pierre}. It is located close to Malargüe, in Argentina, at an average altitude of $\sim \SI{1400}{\m}$. The original triangular grid of surface detectors with a \SI{1500}{\m} spacing is sensitive to cosmic-ray energies above $\sim \SI{e18.5}{\eV}$ ~\cite{ranchon2005response}. Later, two denser arrays with \SI{750}{\m} and \SI{433}{\m} spacings were deployed to extend the energy range of the observatory to $\sim \SI{e16.5}{\eV}$. The underground muon detector consists of scintillators deployed next to the water Cherenkov detectors that constitute these denser arrays~\cite{aab2016pierre}. At each position, three \SI{10}{\m^2} modules are buried at a depth of \SI{2.3}{\m} to shield electromagnetic particles~\cite{aab2021design}.

IceCube, located at the geographic South Pole, is a cubic-kilometer Cherenkov detector, fundamentally designed as a neutrino observatory~\cite{abbasi2013icetop}. Complemented by its surface array IceTop, situated at an altitude of \SI{2835}{\m}, it extends its capabilities to cosmic-ray physics. IceTop consists of a \SI{1}{\km^2} array with 162 water Cherenkov detectors in 81 stations with a \SI{125}{\m} spacing. It measures the electromagnetic component and predominantly low-energy muons from extensive air showers generated by cosmic rays exceeding $\SI{e14}{\eV}$. High-energy muons are detected in coincidence in the in-ice detector IceCube. 

Finally, the KASCADE Experiment was located at the Forschungzentrum Karlsruhe in Germany at an altitude of \SI{110}{\m} a.s.l.~\cite{apel2010kascade, antoni2003cosmic}. It consisted initially of an electromagnetic and muon detector array upgraded to KASCADE-Grande by extending the array area to detect cosmic rays with energies between \SI{e14}{\eV} and \SI{e18}{\eV}. The muon detector component, which is of interest for this work, encompassed 252 detector stations arranged on a rectangular grid with a spacing of \SI{13}{\m} covering an area of \SI{0.04}{\km^2} with plastic scintillators of 90$\times$90$\times$\SI{3}{\cm^3} with \SI{10}{\cm} lead and \SI{4}{\cm} iron shields. 

To shed light on the muon puzzle, improving air shower simulation frameworks and enhancing their technical capabilities to integrate new theoretical models is of utmost importance. In this work, we present advancements in these aspects: in Sec.~\ref{sec:sims}, we introduce a software extension to improve computing times, enabling the massive simulation of ultra-high energy air showers with full detailed 3 dimensional and timing information (3D) for all particles. Then, in Sec.~\ref{sec:EASphysics}, we use this extension to compare muon-related air shower observables for different experiments. Finally, in Sec.~\ref{sec:CC}, we discuss the impact on the muon content of air showers after a theoretical model modification based on the core-corona approach~\cite{baur2023core}. 

\section{Air shower simulation framework: CONEX option in CORSIKA}\label{sec:sims}

The interpretation of cosmic-ray data has historically relied on Monte Carlo simulations that provide a detailed description of the air shower development. As the primary cosmic-ray energy increases, the number of particles in the shower also increases, making these simulations computationally expensive, even with boost techniques such as the \textit{thinning algorithms} THIN and THINMAX, implemented in the CORSIKA framework~\cite{heck1998corsika, hillas1981two}. Here, particles produced in an interaction with energy below a predefined threshold are grouped. From each group, only a randomly selected particle's evolution is simulated and stored. Energy conservation is accounted for by assigning an appropriate weight to the products of this particle (the inverse of the energy fraction taken by this particle). The thinning process is stopped when a maximum weight is reached. The THINMAX algorithm further optimizes computation by putting priority on maximizing the weight over strict energy conservation, thus leading to a lower number of particles but with the same predefined maximum weight. To further improve this situation, the air shower simulation software CONEX employs a hybrid approach~\cite{bergmann2007one}, combining Monte Carlo treatment of high-energy particles (similar to the CORSIKA framework) with a numerical description of low-energy showers based on the solutions of cascade equations~\cite{dedenko1965new, hillas1965calculations, Stanev:1994pf, Drescher:2002cr, bossard2001cosmic}. This hybrid method improves computational efficiency and allows for more simulation repetitions, making CONEX more suitable for testing modifications of hadronic properties.

Describing extensive air showers requires two systems of cascade equations: one for the hadronic component and another for the electromagnetic component. In CONEX's hadronic equations, we consider only protons, neutrons, charged pions, and kaons as \textit{projectile particles}. Projectile-air interactions produce \textit{secondary particles} such as protons, neutrons, pions, kaons, photons, muons, and electrons. Other hadron types produced by decaying particles or through interactions are assumed to decay immediately into some of these secondary particles. Photons and electrons, once created, are transferred directly to the electromagnetic equations.
 
\subsection{CONEX}\label{subsec:conex}

In the CONEX framework, the cascade initiated by the primary hadron is simulated through explicit Monte Carlo until the produced secondaries reach an upper energy threshold. All subthreshold particles are filled into the \textit{source terms} of the cascade equations, which give the initial conditions for the numerical analysis. Then, the hadronic and electromagnetic cascade equations are solved for each depth level. The spectra that result from the $\mathrm{n}^\mathrm{th}$ level feed the source terms of the $\mathrm{(n+1)}^\mathrm{th}$ level. Hadronic and electromagnetic particles are added to their respective source terms, regardless of the shower component from which they originate.  

The hadronic cascade equations take into account all the possible processes in the following system of integro-differential equations:
\begin{align} \label{CE}
\frac{\partial h_a(E,X)}{\partial X} =
      & - \frac{h_a(E,X)}{\lambda_a(E)} - \frac{h_a(E,X)}{\tau_a(E) \; \rho_\mathrm{air}(X)} + \frac{\partial}{\partial E} \; \left( \beta_a^\mathrm{ion}(E) \; h_a(E,X) \right) \nonumber \\
	  & + \sum_d \int_{E}^{E_\mathrm{max}} dE' \; h_d(E',X) \; \left(  \frac{ W_{d \rightarrow a}(E',E)}{\lambda_d(E') } + \frac{D_{d \rightarrow a}(E',E)}{\tau_d(E') \; \rho_\mathrm{air}(X)} \right)  \nonumber \\
	  & + S_a^\mathrm{had}(E,X),
\end{align}
where $h_a(E,X)$ is the differential energy spectrum of hadron type \textit{a}, with energy $E$ at depth position $X$ along a given straight line trajectory. $\beta_a^\mathrm{ion}=-dE_a/dX$ is the ionization energy loss per depth unit, $\lambda_{\textit{a}}=\textit{m}_\mathrm{air}/\sigma^{\textit{a}-\mathrm{air}}_\mathrm{inel}$ the mean free path, and $\tau_a$ the lifetime in the laboratory system (related to the proper lifetime by $\tau_a=\tau_a^{0} E/m_a$). $m_\text{air}$ and $m_a$ are the nuclear mass of air and the mass of hadron type \textit{a}, respectively, while $\sigma^{\textit{a}-\mathrm{air}}_\mathrm{inel}$ is the inelastic cross section of an interaction between this hadron type and an air nucleus. $W_{d \rightarrow a}$ and $D_{d \rightarrow a}$ are the inclusive secondary spectra for interactions and decays, respectively. Muons are treated as hadrons, but do not have interaction terms. The five terms in Eq.~\ref{CE} describe the variation in the hadron number due to interactions with air nuclei, particle decays, ionization loss, hadron production from higher-energy parents (with energies up to $E_\mathrm{max}$) and source terms. The particle decay term results from the decay rates $\text{d}h_a=-h_a \text{d}L/(\tau_a c)$ and $\text{d}L/\text{d}X=\rho_\mathrm{air}^{-1}(X)$, where $c$ is the speed of light and $\rho_\mathrm{air}(X)$ is the air density at depth $X$.

The source term $S_a^\mathrm{had}(E,X)$, defining the initial conditions, has three components:
\begin{equation}
	S_a^\mathrm{had}(E,X)=S_a^{\mathrm{MC} \rightarrow \mathrm{had}}(E,X) + S_a^{\mathrm{em} \rightarrow \mathrm{had}}(E,X) + S_a^{\mathrm{em} \rightarrow \mu}(E,X).
\end{equation} 
The first term consists of all contributions of sub-threshold hadrons produced during the Monte Carlo simulation of above-threshold particles:
\begin{equation}
	S_a^{\mathrm{MC} \rightarrow \mathrm{had}}(E,X)=\sum_i \delta^a_{d_i} \; \delta(E-E_i) \;\delta(X-X_i),
\end{equation}
with $d_i$, $E_i$, $X_i$ being the type, energy, and depth position of the source particle $i$. The second term includes the hadrons that come from photonuclear interactions in the electromagnetic cascade equations. The production distributions $W_{\gamma \rightarrow a}(E',E)$ are approximated at high energy with $\pi^{+/-}$-air interactions and at low energy with $\rho^0$ and $\omega$ resonances: 
\begin{equation} \label{eq:CE-SourceEm2Had}
    S_a^{\mathrm{em} \rightarrow \mathrm{had}}(E,X) = \int_{E}^{E_\mathrm{max}} dE'   \; l_{\gamma}(E',X) \; W_{\gamma \rightarrow a}(E',E) \; \tilde{\sigma}_{\gamma}^{\text{photonuclear}}(E') , 
\end{equation}
where $l_{\gamma}$ is the photon energy spectrum. Finally, the photoproduction of muon pairs gives rise to the last contribution in the hadronic source term:
\begin{equation} \label{eq:CE-SourceEm2Muon}
   S_a^{\mathrm{em} \rightarrow \mu}(E,X) = \int_{E}^{E_\mathrm{max}} dE' \; l_{\gamma}(E',X) \; W_{\gamma \rightarrow \mu}(E',E) \; \tilde{\sigma}_{\gamma}^{\mu-\text{pair}}(E').
\end{equation}

\begin{figure}[t]
    \centering
    \includegraphics[width=1.0\textwidth]{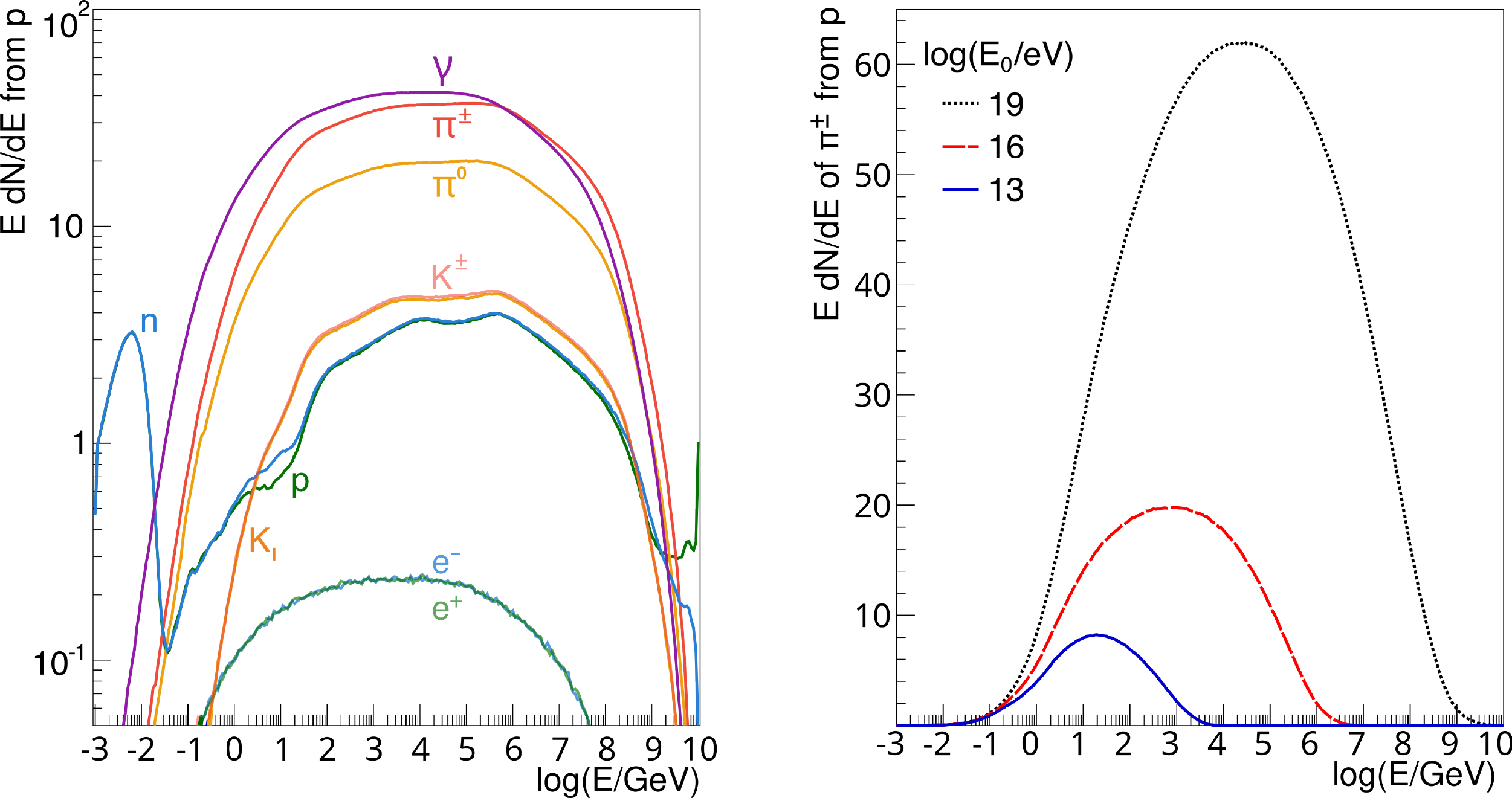}
    \caption{Left: Energy spectra of secondary particles from \textit{p}-air interactions at \SI{e19}{\eV} using \EPOSLHC. Different colors denote different particle species. Right: $\pi^{\pm}$ energy spectra from \textit{p}-air interactions for different proton energies using \QGSJETIId.}
	\label{fig:CONEX}
\end{figure}

Analogously, the electromagnetic equations apply to electrons, positrons, and photons. The equally weighted interaction processes considered are Bremsstrahlung, Bhabha, Moeller, pair annihilation, and pair production with no decay term. This approach allows for a partial analytical solution using eigenvalues in contrast to the numerical solution of the hadronic equations~\cite{bergmann2007one}.

CONEX implements Monte Carlo simulations with the updated high-energy hadronic interaction models \EPOSLHC~\cite{Pierog:2013ria}, \QGSJETIId~\cite{Ostapchenko:2010vb}, and \SIBYLLd~\cite{Riehn:2019jet}. In the numerical analysis, the same models are used to precalculate the spectra of secondary particles $W_{d \rightarrow a}$: for each model, secondary particle type, projectile particle type, and projectile energy, CONEX provides one spectrum. The projectile energy is discretized in 20 logarithmic energy bins per decade from \SI{e9}{\eV} to \SI{e19}{\eV}, which results in $\sim 10^{4}$ secondary particle spectra. For example, Fig.~\ref{fig:CONEX} (left panel) shows all the \EPOSLHC\ energy spectra of secondary particles resulting from an interaction between a \SI{e19}{\eV} proton and an air nucleus. The right panel shows the charged pion energy spectra from proton-air interactions for three different proton energies using \QGSJETIId. Modifying these spectra changes in an effective way the properties of hadronic interactions used in air shower simulations.

\subsection{Extension to multi-dimensional distributions}\label{subsec:CONEX3D}

\begin{SCfigure}[1.2][t]
\centering
\includegraphics[width=0.45\textwidth]{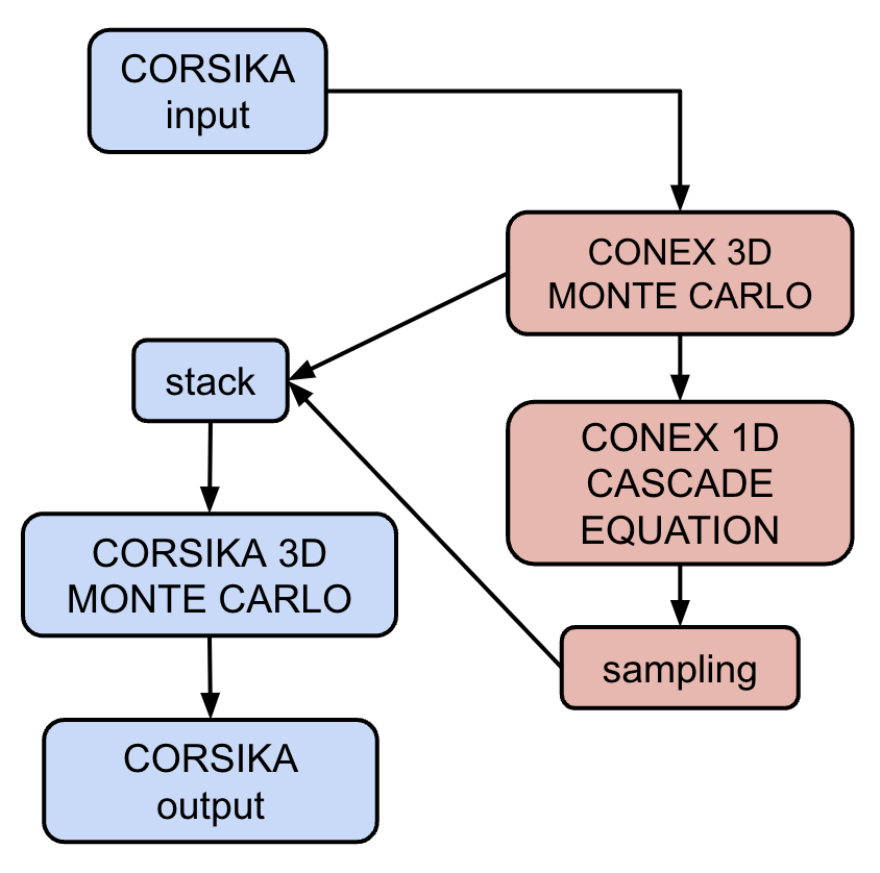}
\caption{Flow chart of the CONEX option in CORSIKA. The CORSIKA structure manages the steering files that set the simulation parameters and transfers these to begin the Monte Carlo simulation in CONEX. CONEX uses cascade equations to handle particles with energy below a predefined threshold. CORSIKA recovers then secondaries from the Monte Carlo or cascade equations with energy below a second threshold and treats them again with 3D Monte Carlo. From the solutions of the cascade equations, individual particles are sampled and saved into the CORSIKA stack.}
\label{fig:diagram}
\end{SCfigure}

CONEX only provides the energy distributions of all particles along the shower axis, allowing thus fast and realistic 1-dimensional simulations of the \textit{longitudinal distribution}, which describes the shower development as a function of atmospheric depth. To obtain 3-dimensional (3D) distributions of particles at the ground, CONEX has been included in the CORSIKA package~\cite{Pierog:2011kzx} as outlined in Fig.~\ref{fig:diagram}. In this integration, the numerical analysis returns to Monte Carlo tracking when particles fall below a second energy threshold~\cite{Drescher:2002cr}. Here, we propagate in 3D the particles generated in the cascade equations that fall below the low-energy threshold, as well as the low-energy particles produced in the initial CONEX Monte Carlo part.

We select the second transition threshold by balancing computation time with accuracy: the threshold must be low enough to reduce the Monte Carlo contribution but high enough to avoid biases. At the highest energies, the longitudinal momentum is orders of magnitude higher than the typical $\lesssim \SI{10}{\giga\eV}$ transverse momentum, and the emission angle along the shower axis is small. Consequently, at a few TeV, we can neglect this angle~\cite{Drescher:2002cr}, and CONEX's cascade equations will accurately describe the shower development. To further reduce this threshold, we compute the mean transverse momentum of the secondary particles using the cascade equations, thus decreasing the transitions to Monte Carlo when emission angles become significant. Since CONEX solves the equations at consecutive depth levels, this procedure results in a (1+1)D way of describing the air shower development.

\begin{figure}[t]
\centering
\includegraphics[width=0.9\textwidth]{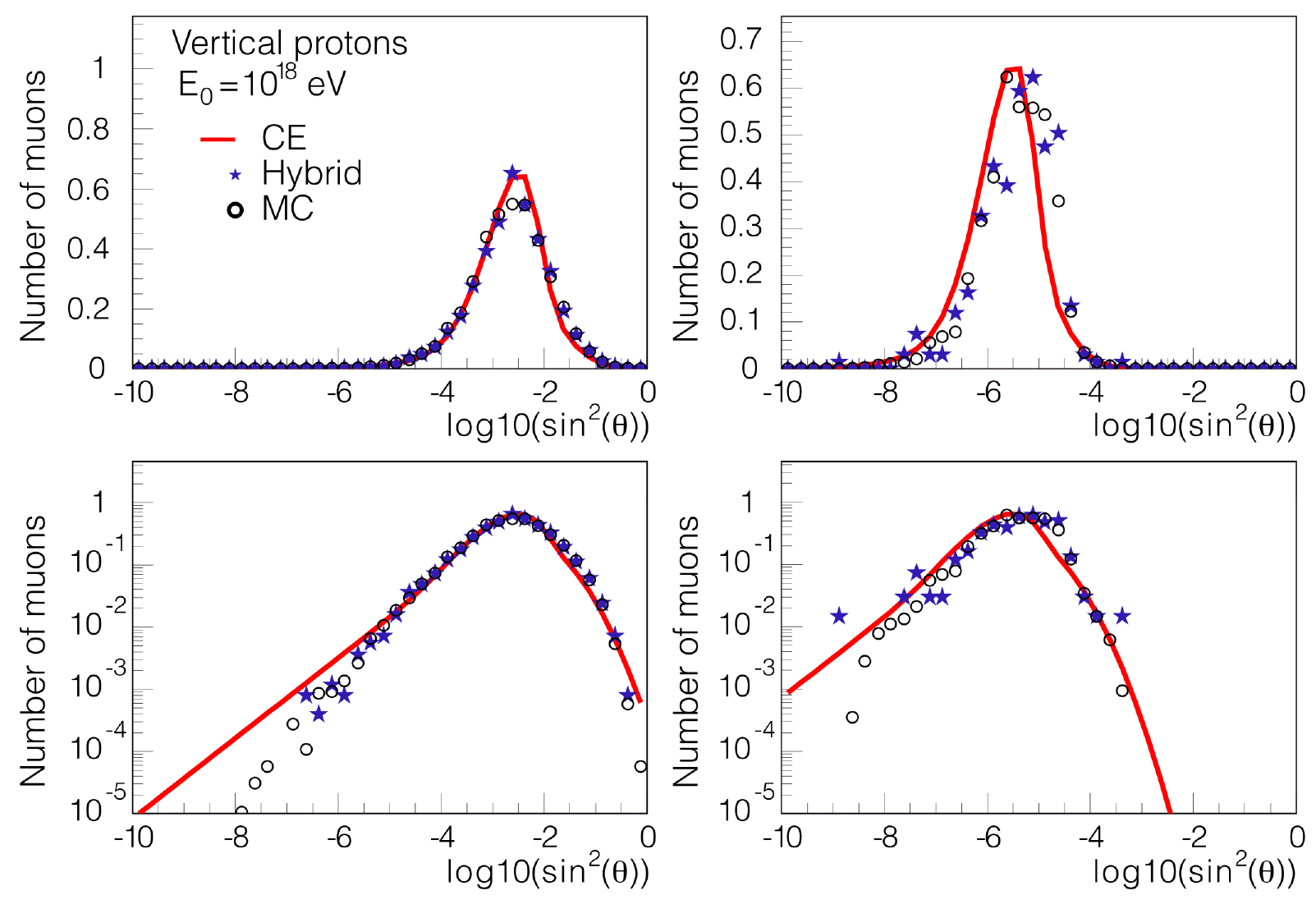}
\caption{Normalized muon angular distributions for a proton-initiated shower with primary energy $E_0=\SI{e18}{\eV}$ using \QGSJET, where $\theta$ is the angle between the muon direction and the shower axis. Red lines represent the distribution obtained from cascade equations, while blue stars are muons sampled from the corresponding cascade equations in CONEX. Finally, open circles are the result of pure Monte Carlo simulations. Top: Linear scale. Bottom: Logarithmic scale. Left: Muons with energy around \SI{5}{\GeV}. Right: Muons with energy around \SI{180}{\GeV}.}
\label{Angular_All_Muons}
\end{figure}

In the cascade equations, we compute the secondary particles' mean transverse momentum assuming a realistic angular distribution instead of a bare longitudinal propagation. To track the properties of each hadronic interaction model, the $\langle p_{\rm t}^2 \rangle$ values are tabulated in $T_{d \rightarrow a}$ spectra ($T^{W}$ for interaction terms and $T^{D}$ for decay terms) for each projectile energy, projectile particle type and hadronic interaction model, as a function of secondary particle type and energy. Thus, we can write another set of cascade equations for hadrons and muons that describe the accumulation of transverse momentum:
\begin{align} \label{CEpt2}
\frac{\partial p_{{\rm t}a}^2(E,X)}{\partial X} =
    & \sum_d \int_{E}^{E_\mathrm{max}} dE' \; h_d(E',X) \; \nonumber \\
    & \left( \frac{ W_{d \rightarrow a}(E',E)}{\lambda_d(E') } \; \left[ T^{W}_{d \rightarrow a}(E',E) + C_a \; (\langle p_{{\rm t}d}^2(E',X) \rangle,E') \right] \right. \nonumber \\
    & \left. + \frac{D_{d \rightarrow a}(E',E)}{\tau_d(E') \; \rho_\mathrm{air}(X)} \; \left[ T^{D}_{d \rightarrow a}(E',E) + C_a \; (\langle p_{{\rm t}d}^2(E',X) \rangle,E') \right]  \right) \nonumber \\
    & + S_a^{p_{\rm t}^2}(E,X).
\end{align}
Using that $\langle p_{\rm t}^2(E,X) \rangle = \sum_h p_{\rm t}^2(E,X) / h(E,X)$, we add the $\langle p_{\rm t}^2 \rangle$ of a previous generation via a correction function $C_a$ for particle type $a$. We use this function to fine-tune $\langle p_{\rm t}^2 \rangle$ in the cascade equations based on Monte Carlo simulations. Of all hadrons, only the last generation makes a significant contribution. Thus, we can neglect the transverse momentum of generations before the last one ($C_a\equiv0$ in those cases). For muons, whose main contributions come from the decay of pions and kaons, we account for the decaying particle's transverse momentum $\langle p_{\rm t}^2 \rangle$ in an effective way via 
\begin{align} \label{CEcorr}
C_\mu(\langle p_{\rm t}^2 \rangle,E)=
    &\max \left( 0.4\;\langle p_{\rm t}^2 \rangle - \frac{ \langle p_{\rm t}^2 \rangle^2}{P_\mu^2}, 0 \right),
\end{align}
where $P_\mu^2=(E+2m)E$ is the total momentum of the muon, since $E=E_{\rm tot}-m$ is its kinetic energy. This formula has been optimized to reproduce the muon angular distribution as obtained from Monte Carlo simulations. The parameter 0.4 reflects the fact that only a fraction of the decaying particle's transverse momentum should be taken into account. The negative term is there to guarantee that the added transverse momentum does not exceed the total momentum of the muon. Particles are then sampled following a (Moyal + Normal) distribution using $\langle\sin^2(\theta)\rangle$ and $\langle\sin^2(\theta)\rangle^2$ as parameters:
\begin{align} \label{CEmoyal}
f(x,\mu,\sigma)=
    &\frac{1}{2} \; \frac{\exp\left[ -\frac{1}{2}\left( \frac{\mu-x}{\sigma}+e^{-\frac{\mu-x}{\sigma}} \right) \right] } {\sigma\sqrt{2\pi}} \nonumber \\
    &+\frac{1}{2} \; \frac{\exp\left[ -\frac{1}{2} \frac{(\mu-x)^2}{2\sigma^2} \right]}{\sigma\sqrt{4\pi}} ,
\end{align}
with $\mu(a,b)=\log_{10}a-0.5\sigma^2(a,b)$ and $\sigma^2(a,b)=0.5\log_{10}(1+b/a^2)$, where $a=\langle\sin^2(\theta)\rangle$, $b=\langle\sin^2(\theta)\rangle^2$, and $\langle\sin^2(\theta)\rangle$ is given by $\langle p_{\rm t}^2 \rangle/P^2$. In Fig.~\ref{Angular_All_Muons}, we show angular distributions obtained through this procedure and from full Monte Carlo simulations, which are consistent.

In this approach, we set a \SI{1}{\tera\eV} low-energy threshold for transferring hadrons from CONEX to CORSIKA's tracking system, ensuring an accurate description of the particle density as a function of distance from the shower core, commonly referred to as the \textit{lateral distribution function}. To track muons in the Earth's magnetic field, we transfer them from the CONEX Monte Carlo directly to the CORSIKA Monte Carlo and systematically sample them back from the cascade equations to the CORSIKA Monte Carlo at each step (every \SI{10}{\g/\cm^2}).

\begin{nolinenumbers}
\begin{table}[b!]
\centering
\renewcommand{\arraystretch}{1.25}
\begin{tabularx}{\textwidth}{l Y  Y  Y} 
  \toprule
    Parameter                      &   KASCADE   & IceTop    & Auger \\
    \midrule
    Primary energy [$\log_{10}$(E/eV)]     &   16        & 17        & 19    \\
    Zenith angle [$^{\circ}$] &   0-27-41-53-67 &  0-27-41-53-67 &  0-27-41-53-67       \\
    Altitude [m]                   &   150       & 2800      & 1450       \\
    Optimal distance from core $d$ [m]     &   100-200   & 400-500   & 1000-1100\\
    Muon energy threshold  $E_{\text{th}}$ [MeV]    &   230       & 200       & 300      \\
    Number of showers              &   240       & 240       & 90   \\
    \bottomrule
    \end{tabularx}
  \caption{Air shower and simulation parameters used in the simulation library for the present work.}
  \label{table:param}
\end{table}
\end{nolinenumbers}

\begin{figure}[t!]
    \centering
    \includegraphics[width=1.0\textwidth]{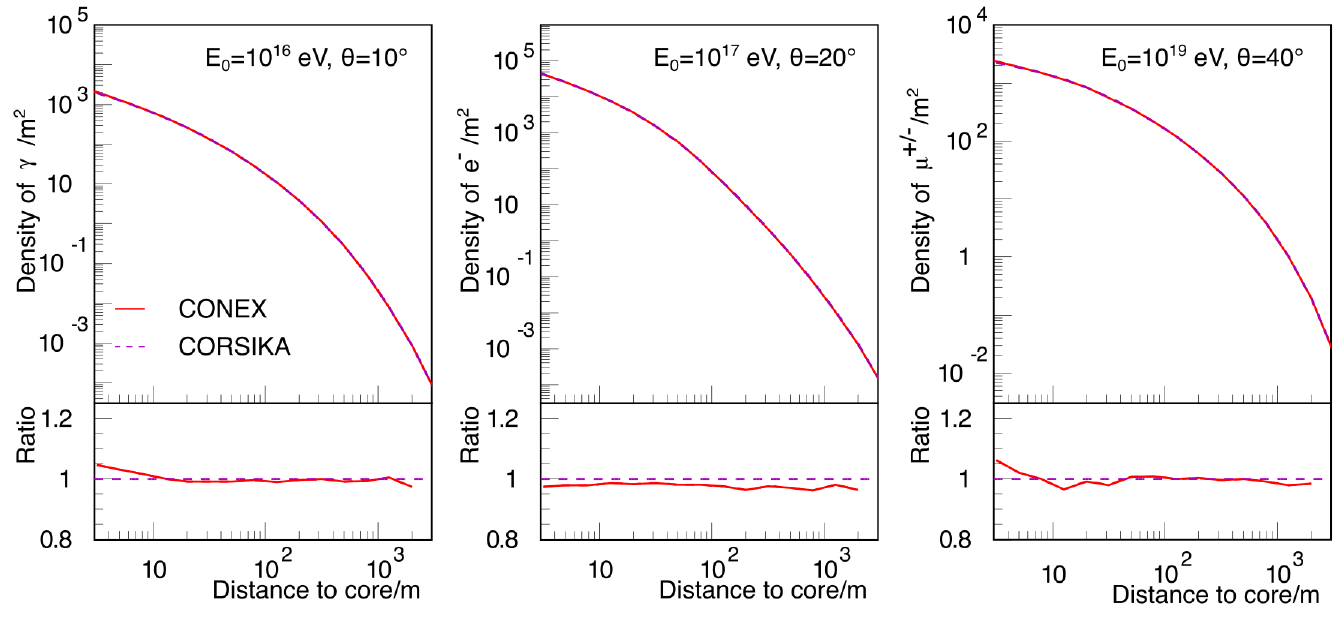}
    \caption{Lateral distribution functions at ground level (top) for proton-initiated showers using CORSIKA and CONEX~3D (simulations based on \EPOSLHC), together with the ratio between models (bottom). Left: photons for a shower with $E_0=\SI{e16}{\eV}$ and $\theta = 15^\circ$ at KASCADE. Middle: electrons for  a shower with $E_0=\SI{e17}{\eV}$ and $\theta = 20^\circ$ at IceTop. Right: muons for  a shower with $E_0=\SI{e19}{\eV}$ and $\theta = 40^\circ$ at Auger.}
    \label{fig:ldf:CONEXvsCORSIKA}
\end{figure} 

\begin{figure}[t] 
    \centering
    \includegraphics[width=1.0\textwidth]{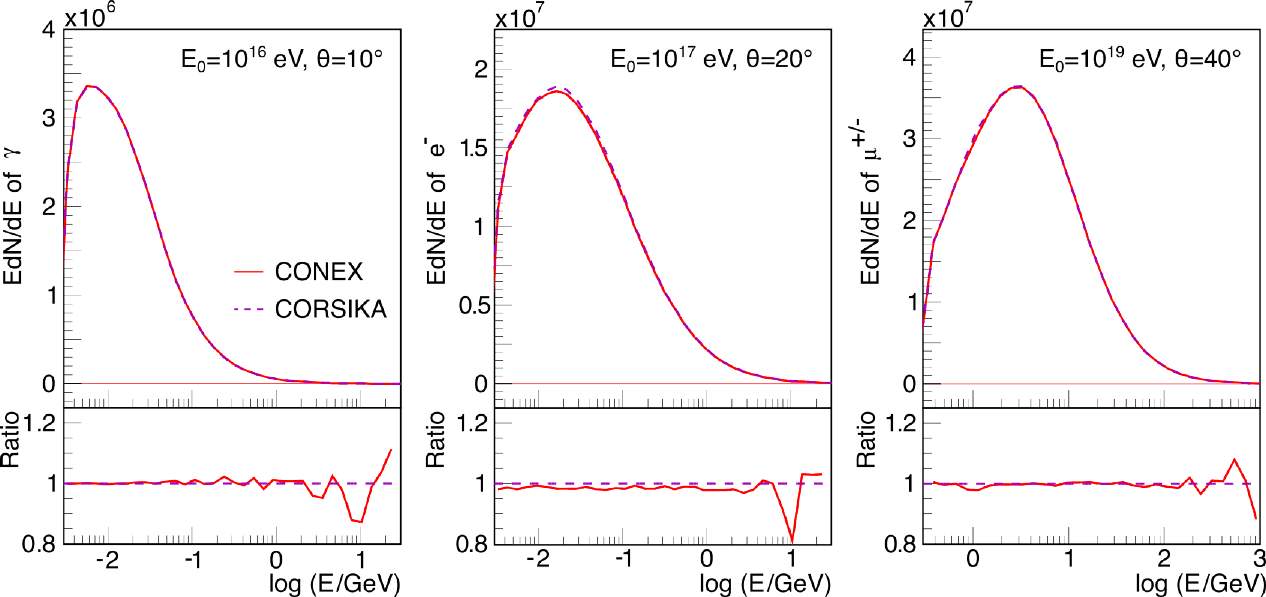}
    \caption{Energy spectra of particles at ground level (top) integrated over all radial distances and obtained with CORSIKA and CONEX~3D for proton-initiated showers (using \EPOSLHC), together with their corresponding ratios (bottom). Left: photons for a shower with $E_0=\SI{e16}{\eV}$ and $\theta = 15^\circ$ at KASCADE. Middle: electrons for  a shower with $E_0=\SI{e17}{\eV}$ and $\theta = 20^\circ$ at IceTop. Right: muons for  a shower with $E_0=\SI{e19}{\eV}$ and $\theta = 40^\circ$ at Auger.}
    \label{fig:Espec:CONEXvsCORSIKA}
\end{figure} 

We sample the electromagnetic particles from the solutions of the cascade equations to be processed by the CORSIKA Monte Carlo following the same procedure. However, since the Coulomb interaction determines the scattering angle, we decompose this potential into higher-order moments and analytically compute the resulting cascade equations at each slant depth step. The second moment allows the computation of $\langle\cos^2(\theta)\rangle$, from which we extract $\langle\sin^2(\theta)\rangle=1-\langle\cos^2(\theta)\rangle$ (needed as argument in Eq.~\ref{CEmoyal}), leading to an increase of the short computational time for the cascade equation resolution by only a factor of two~\cite{Chernatkin:2003rm,Chernatkin:2005}. We set the transition threshold for electromagnetic particles to a kinetic energy of \SI{10}{\giga\eV}. While muons and hadrons can propagate large distances, electromagnetic particles are strongly attenuated. To avoid sampling particles absorbed before reaching the ground, we define the minimum vertical depth (or height above the observation level) below which we track electromagnetic particles. Particles remain in the cascade equations above this cut, by default \SI{400}{\g/\cm^2}, but can still produce low-energy hadrons or muons. 

To validate the new framework, we compare CORSIKA and CONEX~3D simulations for the three configurations representative of the main current experiments. We summarize in table~\ref{table:param} the respective cosmic ray primary energy and zenith angles considered, together with the altitude above sea level, the muon energy threshold, and the optimal distance to the shower core appropriate for each experiment. We selected the muon energy threshold in KASCADE based on the minimum energy required to penetrate the detector shield~\cite{apel2010kascade}. For Auger and IceTop, we determined the muon energy threshold by considering their energy distributions and experiment altitudes, ensuring it is sufficiently low to capture most particles (see Fig.~\ref{fig:SpectraDistances}). Similarly, for electromagnetic particles, we used a threshold of $E_{\text{th}}=\SI{3}{\mega\eV}$.

\begin{figure}[t]
    \centering
   \includegraphics[width=1.0\textwidth]{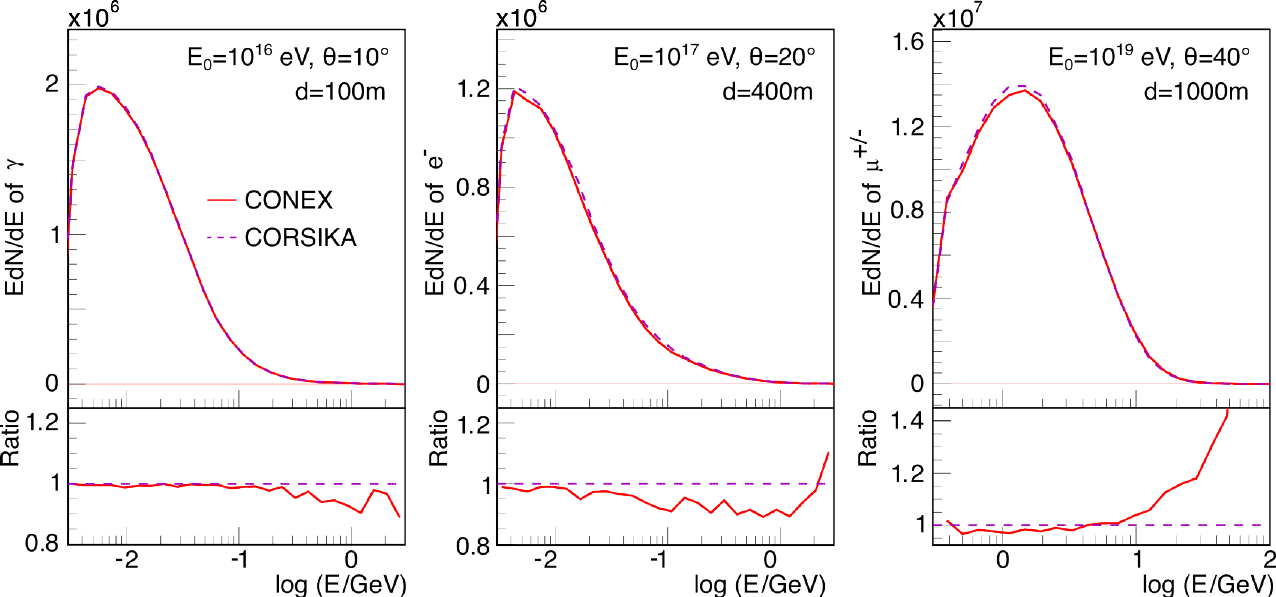}
    \caption{Energy spectra for proton showers of particles at ground level and at each experiment's optimal distance (top) obtained with CORSIKA and CONEX~3D (using \EPOSLHC), and their corresponding ratios (bottom). Left: photons for a shower with $E_0=\SI{e16}{\eV}$ and $\theta = 15^\circ$ at KASCADE. Middle: electrons for a shower with $E_0=\SI{e17}{\eV}$ and $\theta = 20^\circ$ at IceTop. Right: muons for  a shower with $E_0=\SI{e19}{\eV}$ and $\theta = 40^\circ$ at Auger.}
    \label{fig:CONEXvsCORSIKA}
\end{figure}

Figs.~\ref{fig:ldf:CONEXvsCORSIKA},\ref{fig:Espec:CONEXvsCORSIKA}, and \ref{fig:CONEXvsCORSIKA} display the mean lateral distribution functions, the energy spectra over all distances from the shower core, and the energy spectra at the optimal distance from the core, respectively. In each, the left panel corresponds to the photon distribution ($\gamma$) at the ground for the KASCADE experiment, the middle panel to electrons in IceTop, and the right panel to muons at the Pierre Auger Observatory. Dashed lines labeled ``CORSIKA'' refer to simulations of single-proton initiated showers with only Monte Carlo, and solid lines labeled ``CONEX'' refer to simulations with cascade equations at intermediate energy (as described in Fig.~\ref{fig:diagram}). Since we use the same seeds for the Monte Carlo at high energy, the showers have the same global evolution (same $X_\mathrm{max}$) and can be directly compared at ground level.

In Fig.~\ref{fig:ldf:CONEXvsCORSIKA}, the mean ratio of the muon lateral distribution functions is close to one, with minor discrepancies ($<5\%$) at less than \SI{10}{\m} from the shower core, a distance not relevant for most of the high-energy shower experiments, or in IceTop, that measures showers close to their maximum. We also note an overall slight underestimation of electromagnetic particles near the shower maximum (IceTop), which is limited to a percent level. CONEX accurately reproduces the muon energy spectra across the entire energy range, as demonstrated in Fig.~\ref{fig:Espec:CONEXvsCORSIKA}. At larger distances from the shower core (\SI{400}{\m} and \SI{1000}{\m}), as shown in Fig.~\ref{fig:CONEXvsCORSIKA}, the energy spectra exhibit more significant discrepancies, particularly an excess of high-energy muons in CONEX simulations. Nevertheless, the peak is well-reproduced in each case, and the high-energy tail has a minimal impact on experimental results. Photons, electrons, and muons behave equally for the three tested phase spaces.

\begin{nolinenumbers}
\begin{table}[b]
\centering
\renewcommand{\arraystretch}{1.25}

\begin{tabularx}{1.0\textwidth}{ l Y Y Y Y Y Y Y   }
\toprule
    
    \multirow{2}{*}{} &  \multicolumn{2}{c}{KASCADE}    & \multicolumn{2}{c}{IceTop}    & \multicolumn{2}{c}{Auger}  \\
                    &   [min] & [MB]     &   [min] & [MB] &   [min] & [MB]    \\
    \midrule
    CORSIKA+THIN  &   35  &  35       &   40  &  304   &   237  &  732    \\
   CORSIKA+THINMAX  &   21  &  20       &   25  &  160   &   210  &  620    \\
    CONEX~3D   &   14   &  24       &   24  &  178    &   22 &  95   \\
    \bottomrule
  \end{tabularx}
    \caption{Computing times (in minutes) and file sizes (in megabytes) for one proton air shower simulation performed with CORSIKA, using the THIN, THINMAX or CONEX~3D options, considering the detector configurations of the KASCADE Experiment ($E_0 = \SI{e16}{\eV}$, $\theta = 15^\circ$), the IceTop Experiment ($E_0 = \SI{e17}{\eV}$, $\theta = 20^\circ$) and the Pierre Auger Observatory ($E_0 = \SI{e19}{\eV}$, $\theta = 40^\circ$). Optimal thinning (see text) and the hadronic interaction model \SIBYLLd\ are used here.}
  \label{table:computingtimes}
\end{table}
\end{nolinenumbers}

In Table~\ref{table:computingtimes}, we summarize the computing times and file sizes for CORSIKA+THIN~\cite{heckextensive}, CORSIKA+THINMAX~\cite{heckextensive}, and CONEX~3D simulations, using the \SIBYLLd\ hadronic interaction model. In the CORSIKA simulations, we use an \textit{optimal thinning} with the following parameters: a thinning threshold given by the fraction $E_{\rm thr}=10^{-6}$ for electromagnetic particles, a maximum weight defined as $W_\mathrm{max}=E_0 \times E_{\rm thr}$ with 100 times lower values for hadrons and muons. We also tested the THINMAX option but it applies only to electromagnetic particles. With this option, all electromagnetic particles reach $W_\mathrm{max}$ exactly, thus improving the computation time and disk occupancy by about $30\%$ and $50\%$, respectively, at the expense of allowing for energy conservation violation when the maximum weight is reached in a given interaction. During particle selection, in some cases energy is gained, while in other cases it is lost. Thus, in average, the total energy is approximately conserved, but not exactly as in the standard thinning option. The thinning process cannot conserve the number of particles and the energy at the same time, while the CONEX 3D option does both. In the Auger case, this improvement is less significant because the simulation spends more time calculating the hadronic shower, while only the electromagnetic shower is affected by THINMAX.

In the CONEX~3D option, we use the same value $W_\mathrm{max}$ to sample the particles from the cascade equations. This results in a very peaked weight distribution at the maximum weight, both for electromagnetic particles and muons. Consequently, there are fewer particles tracked in the Monte Carlo. Thus, the simulation time and disk space can be reduced from a factor of 2 at low energy to a factor of 10 at high energy, while preserving the accuracy of the results. At large distances from the core, where there are fewer particles, this can introduce large statistical fluctuations. We can reach the statistics needed while keeping reasonable simulation times by reducing $W_\mathrm{max}$ for muons and hadrons. As an example, in the Auger case, if we reduce the value of $W_\mathrm{max}$ (in CONEX) for electromagnetic particles and hadrons by a factor of 10, the computation time is still lower than for CORSIKA by about $30\%$, while the statistics at the ground is much larger (leading to smaller artificial fluctuations). In contrast, there is no considerable difference between CONEX~3D and THINMAX simulations at IceTop, since $W_\mathrm{max}=1$ for hadrons and muons. Then, the propagation stops at the maximum of the shower, and using the cascade equations does not reduce the computing time. To conclude, the CONEX~3D option is particularly convenient for high-energy or inclined showers.

\subsection{Simulation library}\label{subsec:library}

The simulation library used in this work consists of three groups representing the main air shower experiments at each altitude (see Sec.~\ref{sec:experiments}). For each group, we run simulations using parameters specific to the corresponding setting. The CONEX parameters aim at avoiding 3D Monte Carlo simulations at high energies and produce ``mean'' showers without fluctuations from the first interaction. This also means that iron induced showers are the superposition of 56 proton showers with a primary energy 56 times lower (superposition model), since the cascade equations cannot deal with nuclei. To increase the statistics in phase space regions with a low number of particles (high energies and large distances from the shower core), we run several simulations with the same parameters, which result in the same mean shower characteristics but with fluctuations due to the randomness in the sampling step (see explanation to Fig.~\ref{fig:diagram}). In this study, we present results for average showers over the whole corresponding set of simulations.

The Auger set comprises 90 mean proton and iron air showers, while IceTop's and KASCADE's have 240. For each set, we simulate five different zenith angles with the same weight ($\sin{\theta}$): $0^{\circ}$, $27^{\circ}$, $41^{\circ}$, $53^{\circ}$ and $67^{\circ}$. We use the shower plane to count particles (projection from the observation plane to the plane perpendicular to the core propagation) and analyze the distance ranges of $d=1000-\SI{1100}{\m}$ for Auger, $d=400-\SI{500}{\m}$ for IceTop and $d=100-\SI{200}{\m}$ for KASCADE. We perform the simulations using CORSIKA's version v77410, which has CONEX v7.5 implemented. In addition, we use the latest hadronic interaction models in each set: \EPOSLHC, \QGSJETIId, and \SIBYLLd. In Table~\ref{table:param}, we summarize the simulation parameters for each air shower experiment and the number of simulations for each set along with the muon energy thresholds and the $d$ values.

\section{Muon-related observables with CONEX}\label{sec:EASphysics}

\begin{figure}[b]
\centering
\includegraphics[width=0.8\textwidth]{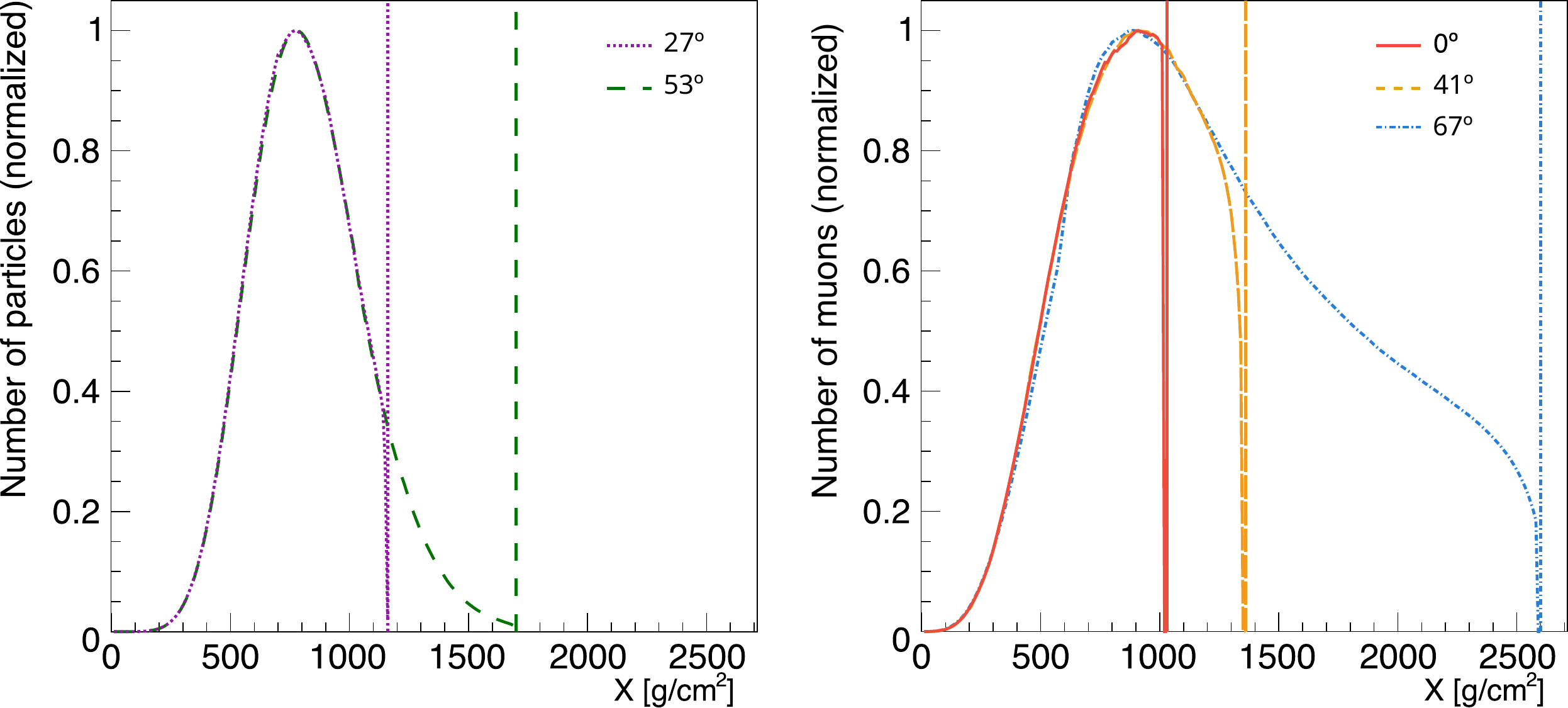}
\caption{Normalized longitudinal profiles of all particles (left) and muons only (right) for proton showers with a primary energy of $E_0=\SI{e18}{\eV}$ and different zenith angles. We used \EPOSLHC\ as the hadronic interaction model. The vertical lines show the position where a shower with the corresponding zenith angle would hit the ground.}
\label{LongProf_All_Muons}
\end{figure}

\begin{figure}[t]
    \centering \includegraphics[width=1.0\textwidth, trim={0 40.5cm 0 0},clip]{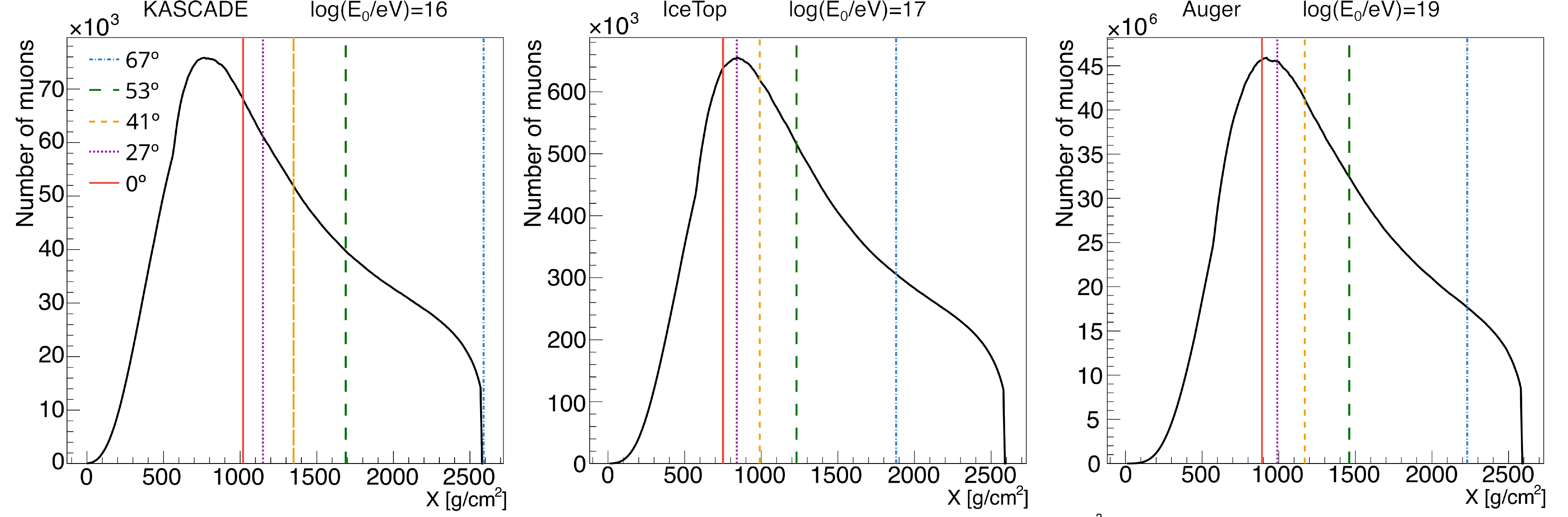}
    \caption{Muon longitudinal distributions for proton-initiated showers with an incidence angle of $67^{\circ}$ using \EPOSLHC. We consider the altitudes and optimal primary energies of KASCADE (Left), IceTop (middle), and Auger (right). The vertical lines show the position of the ground at each site for different zenith angles.}   \label{fig:ObsComp3expNmu}
\end{figure}

\begin{figure}[b]
    \centering
\includegraphics[width=1.0\textwidth, trim={0 20.4cm 0 20.5cm},clip]{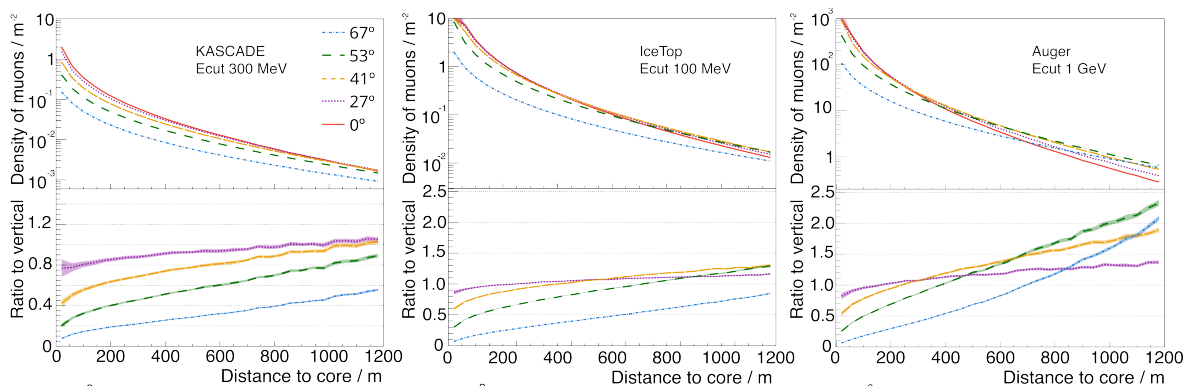}
    \caption{Muon lateral distributions for proton-initiated showers with different incidence angles and at different sites (left: KASCADE, middle: IceTop, right: Auger) using \EPOSLHC. We also display, as a reminder, the muon energy thresholds used (denoted as ``Ecut'').}
    \label{fig:ObsComp3expLDF}
\end{figure} 

To better understand the behavior of muons in extensive air showers, we analyze the number of particles in the shower as a function of traversed overburden, referred to as the longitudinal profile. In Fig.~\ref{LongProf_All_Muons}, we present longitudinal profiles of all particles (left panel) and muons only (right panel), normalized to a maximum value of 1, for proton showers with different incidence angles and primary energy $E_0 = \SI{e18}{\eV}$, using \EPOSLHC\ for the high-energy interactions. Electromagnetic particles dominate the all-particle distributions but have a very strong attenuation after the shower maximum. Thus, after around \SI{1500}{\g/\cm^2} (corresponding to the sea-level for a zenith angle of $55^{\circ}$), this component becomes negligible (it then only stems from the decay of muons into electrons and positrons). In contrast, muon profiles have a slow attenuation, allowing their detection at ground even for very inclined showers. Another effect is observed in the muon profiles: a fast but smooth drop just before reaching the ground. Particles far from the core tend to reach the ground before those near the core, since we analyze particles at the shower plane, and thus are accounted for in the final profile at smaller depths than those near the core. For the same reason, the more inclined the shower is the earlier this effect sets in.

We show in Fig.~\ref{fig:ObsComp3expNmu} the number of muons as a function of shower depth for proton showers at $67^{\circ}$, indicating with vertical lines the depths at which the showers would hit the ground for different primary incidence angles and at experimental sites (different heights). The left panel represents results for the KASCADE scenario, the middle for IceTop, and the right for Auger. Each experiment measures different stages of shower development. KASCADE detects showers for all zenith angles after $N_{\mu}^\mathrm{max}$, while at Auger vertical showers reach the maximum number of muons close to the ground. IceTop can detect showers before and after the value of $N_{\mu}^\mathrm{max}$ is attained, depending on the inclination of the shower.

\begin{figure}[t]
\centering\includegraphics[width=0.9\textwidth]{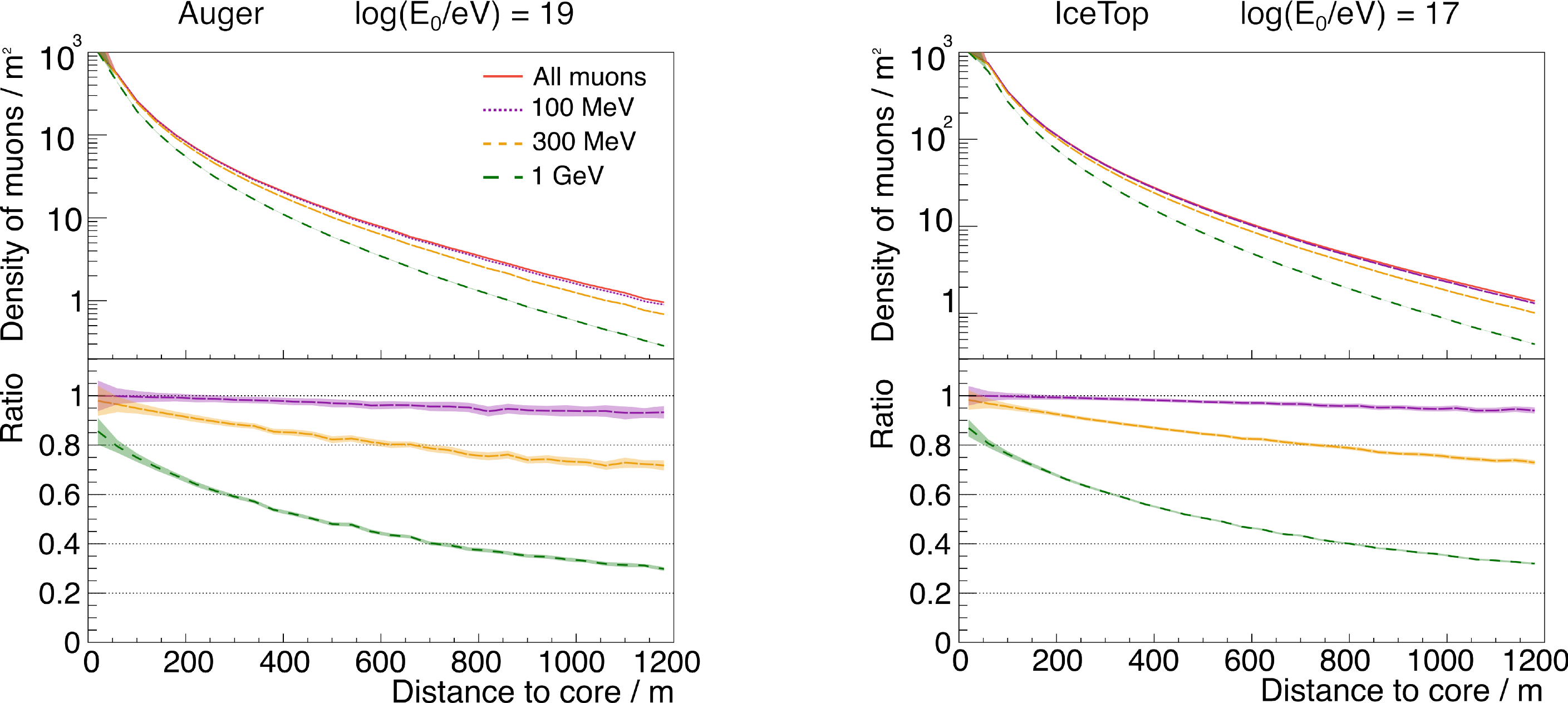}
    \caption{Muon lateral distribution functions for proton-initiated showers at the Pierre Auger Observatory (left) and the IceTop array (right), using \EPOSLHC. We show results for simulations of vertical showers ($\theta=0^\circ$), considering different muon energy thresholds, and the ratio between the lateral distributions for muons with energy above the threshold and all muons.}
    \label{fig:LDFwEth}
\end{figure} 

In Fig.~\ref{fig:ObsComp3expLDF}, we present the muon lateral distributions (top) for different experiments and zenith angles, and the ratio between muon densities for inclined incidence angles and vertical showers (bottom). As the zenith angle increases, the muon density tends to decrease due to the attenuation caused by the increased amount of matter traversed. Concurrently, the distance from the points of production of the particles along the axis to the ground increases with the incidence angle, leading to a flatter shower front. Consequently, the muon density ratio between inclined and vertical showers increases with the distance from the shower core. In the KASCADE scenario, for example, the muon density of showers with $\theta=27^\circ$ is higher than for vertical showers beyond $\SI{900}{\m}$. Note that for IceTop, the muon density for vertical showers is most significant only close to the shower core. This effect originates from the higher altitude of the array (see Fig.~\ref{fig:ObsComp3expNmu}). In this case, we observe an earlier stage of the shower development, and low-energy muons with larger scattering angles are not yet attenuated, contributing to higher ratios compared to KASCADE.

\begin{figure}[b]\centering\includegraphics[width=1.0\textwidth, trim={0 0 0 40.66cm},clip]{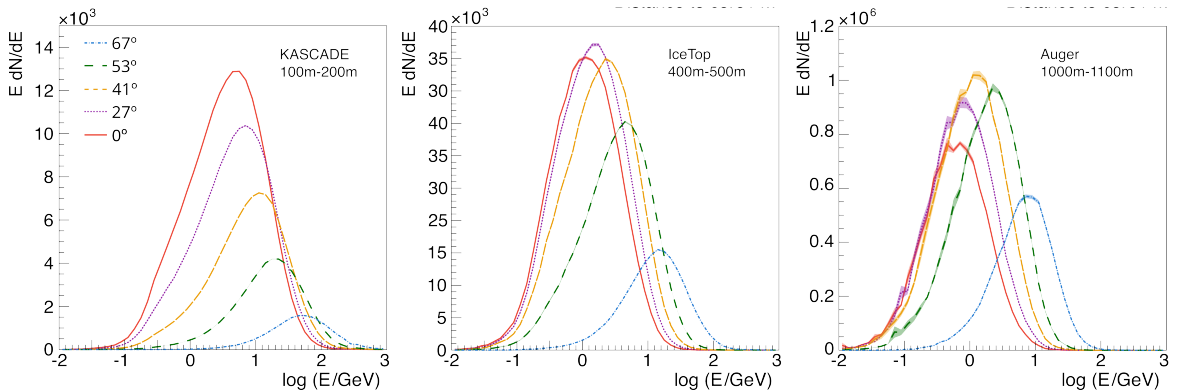}
    \caption{Muon energy spectra for proton-initiated showers with different incidence angles using \EPOSLHC. The optimal distances for KASCADE  (left),  IceTop (middle), and Auger (right) are considered.}\label{fig:ObsComp3expMuonSpectra}
\end{figure} 

\begin{figure}[t]\centering\includegraphics[width=1.0\textwidth]{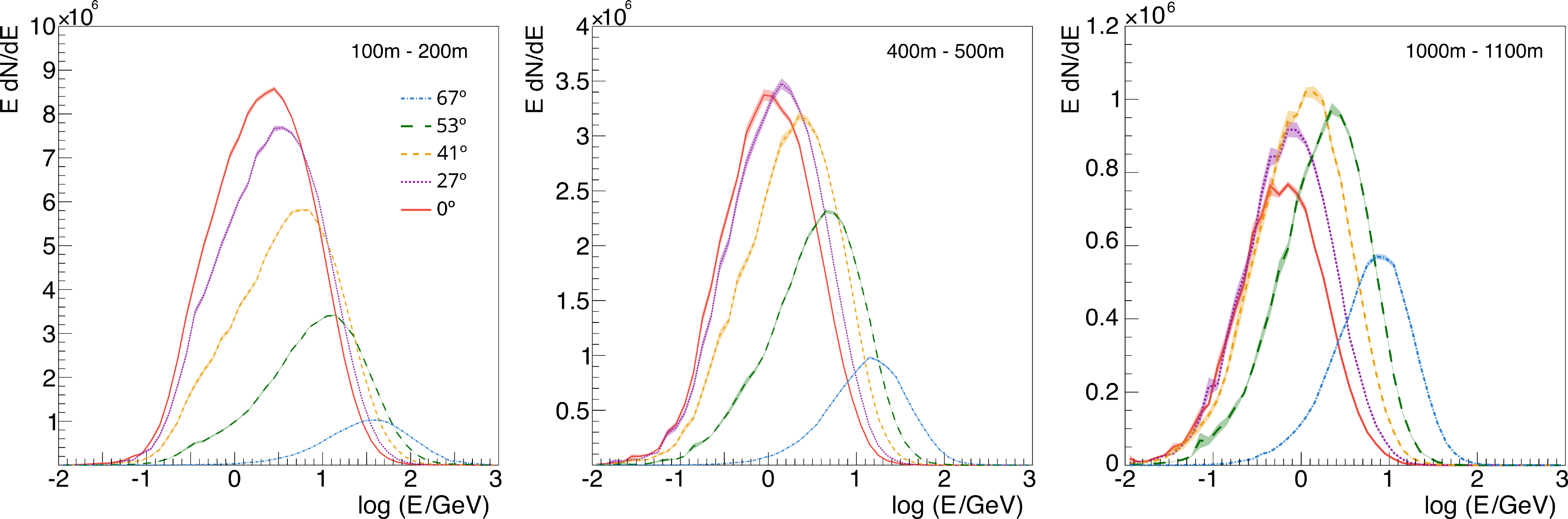}
    \caption{Muon energy spectra at different distances from the shower core and for different zenith angles, as predicted from \EPOSLHC{} simulations of \SI{e19}{\eV} proton showers at the Pierre Auger Observatory.}\label{fig:SpectraDistances}
\end{figure} 

To further understand the differences between the muon lateral distribution functions, we present in Fig.~\ref{fig:LDFwEth} the muon density as a function of the distance to the shower core for proton-initiated vertical showers considering different muon energy thresholds, for the Auger (left) and IceTop (right) scenarios, along with the ratio between the lateral distributions for muons with energy above the threshold and all muons. The similarity between the results for Auger and IceTop highlights that the choice of the muon energy threshold dominates the lateral distribution shaping in Fig.~\ref{fig:ObsComp3expLDF} over other differences between the experiments' properties.

In Fig.~\ref{fig:ObsComp3expMuonSpectra}, we present the muon energy spectra at a characteristic distance from the shower core (see Sec.~\ref{subsec:library}) for each experiment. The position of the peak shifts to higher energies as the angle of incidence increases, mainly because the atmosphere attenuates lower-energy muons. In addition, since inclined showers develop in a less dense atmosphere, the distance between interaction points is larger than the decay length at advanced shower stages. Then, the critical energy at which kaons and pions decay into muons increases. Consequently, the energies of the muons ``shift'' to higher values as the incidence angle increases. 

Furthermore, since KASCADE measures showers after $N_{\mu}^\mathrm{max}$, the number of particles (or height of the distribution peak) decreases with the zenith angle, again as a result of attenuation. However, this behavior is not present in Auger and IceTop. To better understand these differences, we present in  Fig.~\ref{fig:SpectraDistances} the energy spectra for the same set of simulations (\SI{e19}{\eV} proton showers simulated with \EPOSLHC\, at Auger) at the radial distances of 100-$\SI{200}{\m}$ (left), 400-$\SI{500}{\m}$ (center) and 1000-$\SI{1100}{\m}$ (right). Since spectra for different experiments at the same radial distance are similar, the leading effect on the shaping of the muon energy spectra comes from the choice of the distance from the shower core where the shower is measured. As already said, the lateral spread of the muons increases with the incidence angle, causing the position of the peak of the number of muons to be deeper with the increase of the radial distance. As a consequence, depending on the incidence angle, IceTop and Auger may measure shower stages before or after $N_{\mu}^\mathrm{max}$, thus giving a non monotonic behavior of the muon number with respect to the zenith angle.

\begin{figure}[t]
    \centering
    \includegraphics[width=0.5\linewidth]{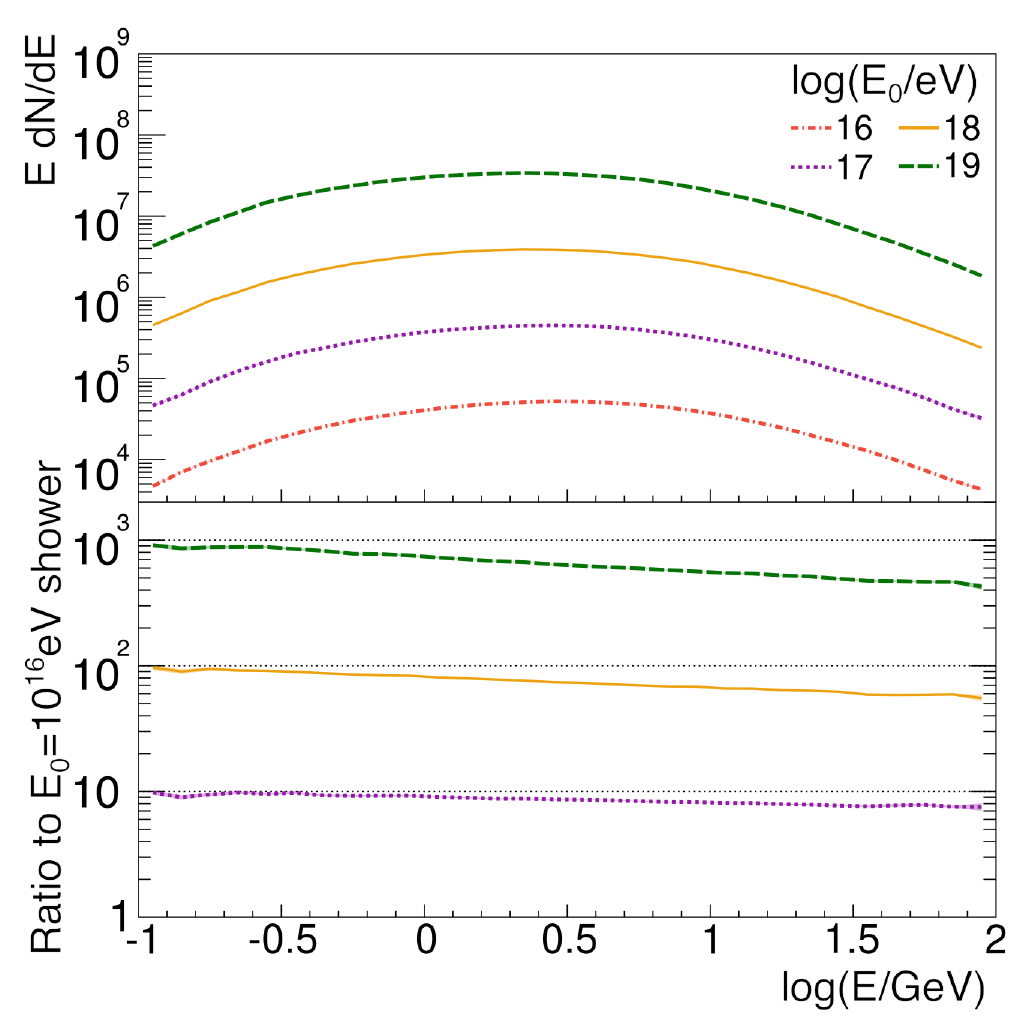}
    \caption{Muon energy spectra for different proton shower energies (top) and the corresponding ratio to the lowest energy (bottom), using \EPOSLHC{} simulations. We show showers at the Pierre Auger Observatory and with an incidence angle of $41^\circ$.}
    \label{fig:test1}
\end{figure}

Finally, Fig.~\ref{fig:test1} shows the simulated muon energy spectra for proton showers at the Pierre Auger Observatory with an incidence angle of $41^\circ$ and different primary energies. The ratios between the energy spectra at a given energy and \SI{e16}{\eV} are almost linear with respect to the muon energy, with a small negative slope. At high energies, muons are mainly the decay products of mesons created in hadronic interactions. This number of muons can be modeled as $(E_0/\xi_c^\pi)^\beta$~\cite{matthews2005heitler}, where $E_0$ is the primary energy, $\xi_c^\pi$ is the pion critical energy and $\beta$ is a parameter with a value close to $0.9$~\cite{engel2011extensive}. This model accurately describes the factor of about $10^\beta$ between consecutive spectra at high muon energies. At lower muon energies, the photonuclear effect plays a dominant role~\cite{muller2019direct}. Here, a photon interacts with a nucleus in the atmosphere and creates some hadrons that subsequently decay into a muon. Since the number of photons is directly proportional to the primary energy, the number of muons resulting from the photonuclear effect is linearly dependent on the primary energy, too (the ratio here is 10 for a 10-fold increase of the energy). In summary, the linear dependence on energy at low muon energies ultimately scales proportionally to $E_0^\beta$, at high energies. For this reason, the gap between spectra decreases as the muon energy increases.

Since the number of muons in air showers constitutes a parameter that is sensitive to the primary cosmic-ray composition, simulations are crucial to determine the relation between muon density and cosmic-ray composition and then interpret the muon content in data. At each experiment, the ratio between the number of muons for iron- and proton-induced air showers depends on the incidence angle, distance to the shower core, and muon energy. 

\begin{figure}[t]
    \centering
    \includegraphics[width=1.0\textwidth, trim={0 0 21cm 0},clip]{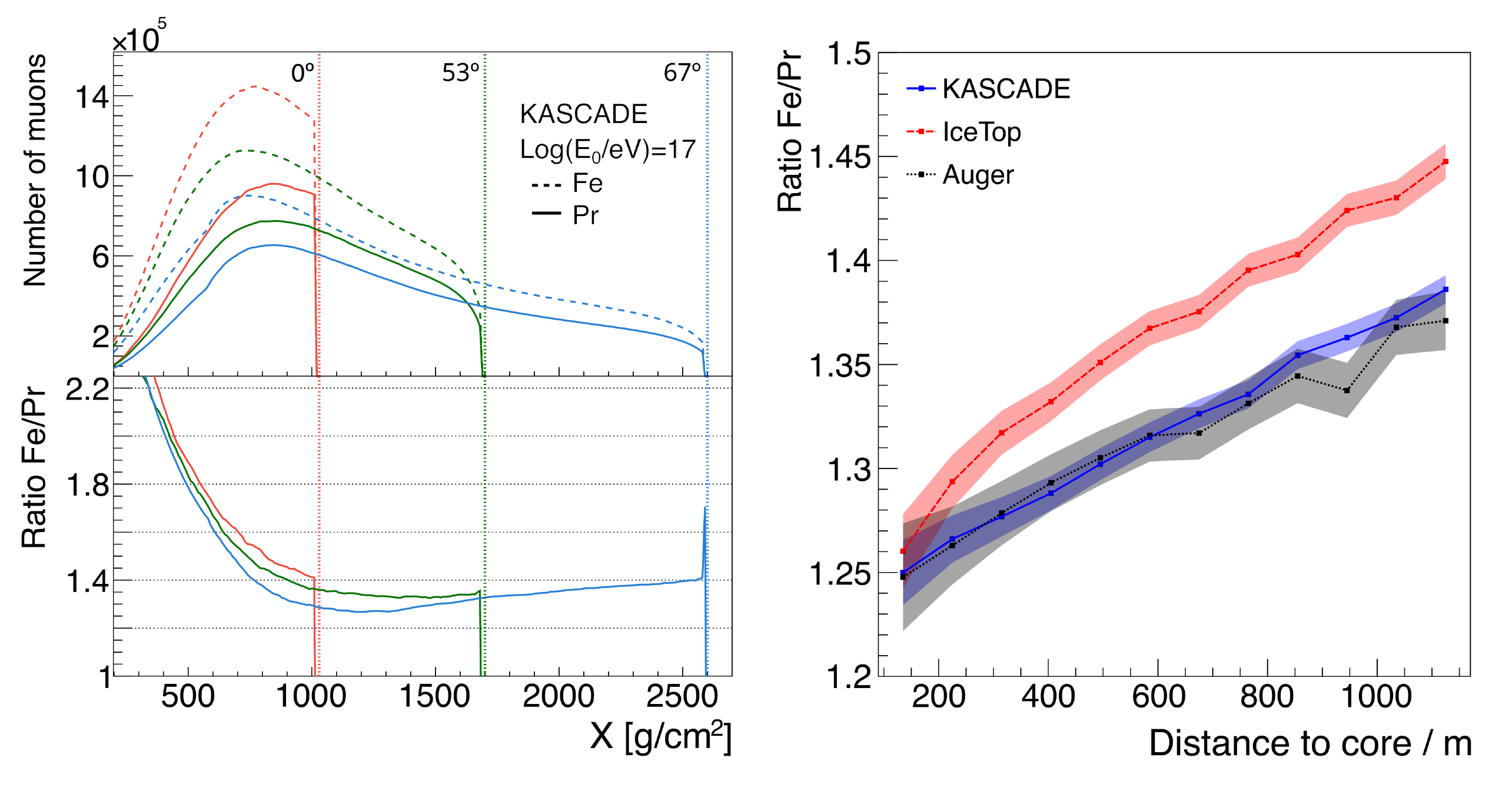}
    \caption{Left: Longitudinal profiles of muons for proton- and iron-initiated showers (top) and the corresponding iron-to-proton ratios (bottom) for showers of \SI{e17}{\eV} primary energy at various incidence angles, as predicted for KASCADE. Right: Ratios of muon lateral distribution functions for iron and proton showers from different experiments, averaged over all zenith angles and weighted by the respective solid angles. Both figures are based on simulations using \EPOSLHC.}
    \label{fig:Long-FePr}
\end{figure} 

\begin{figure}[t]
	\centering
		\begin{subfigure}{0.49\textwidth}
		\centering
		\includegraphics[trim={43.0cm 0.0cm 0cm 0.0cm},clip,width=1.0\linewidth]{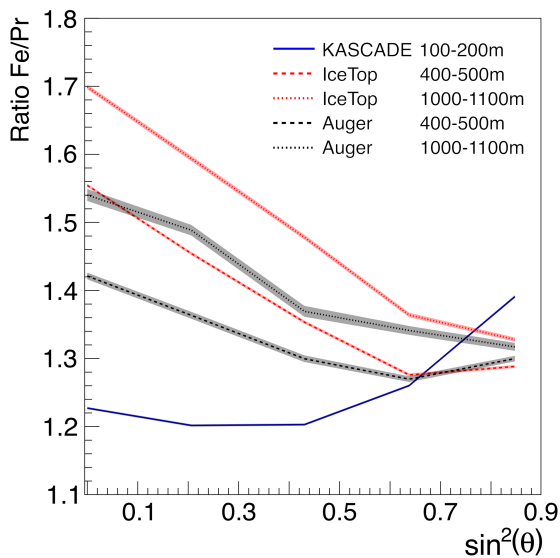}
	\end{subfigure}
	\caption{Ratio of the number of muons between iron and proton showers with a primary energy of \SI{e17}{\eV} (simulated using \EPOSLHC), as a function of the incidence angle, at different experiments and distances from the shower core.}
    \label{fig:FePr_LDF_Spectrum}
\end{figure} 

In the left panel of Fig.~\ref{fig:Long-FePr}, we show the longitudinal profiles of \SI{e17}{\eV} proton (solid lines) and iron (dashed lines) showers, as simulated for KASCADE, and considering different incidence angles, along with the corresponding iron-to-proton ratios. For a fixed experiment altitude, different stages of the shower development are measured (represented by vertical dotted lines in the figure), depending on the incidence angle, which leads to different ratios between proton and iron. 

In the right panel of Fig.~\ref{fig:Long-FePr}, we present the ratio of \EPOSLHC\ muon lateral distribution functions for iron and proton showers, averaged over all zenith angles (with weights that correspond to the solid angle), for the three experiments. The increase in the ratio with the distance to the shower core is due to a geometric effect: iron showers develop higher in the atmosphere and produce more muons further away from the core compared to proton showers at a given altitude. As the distance from the core increases, the muons that are detected originate from higher altitudes.

Finally, we assess the impact of both geometry effects in Fig.~\ref{fig:FePr_LDF_Spectrum}, where we display the ratio of the number of muons as a function of the incidence angle for each experiment and at different distances to the shower core. Generally, the ratio increases when measuring at higher altitudes, further from the shower core, and at smaller incidence angles. The ratio decrease for large zenith angles at Auger and IceCube results from measuring stages close to the shower maximum and muons of lower energy far from the shower core (see Fig.~\ref{fig:SpectraDistances}). The ratio increase for large zenith angles at KASCADE results from measuring higher muon energies (see Fig.~\ref{fig:ObsComp3expMuonSpectra}, left). In addition, since iron showers develop in a less dense atmosphere, pions decay at higher energies, producing more high-energy muons than proton showers. Consequently, the ratio increases weakly with muon energy: in Auger, for example, the ratio is 1.15 for muons with $\SI{0.1}{\GeV}$ and 1.7 for muons of $\SI{10}{\GeV}$.

\section{Simplified core-corona model}\label{sec:CC}

CONEX, either in its stand-alone version or as an option in CORSIKA, implements the latest high-energy hadronic interaction models: \EPOSLHC, \QGSJETIId, and\\ \SIBYLLd. The production of hadrons in these event generators is based on string fragmentation models and fails to reproduce muon-related observations (see Sec.~\ref{sec:experiments}). Recent measurements at the LHC hint towards the existence of other production mechanisms, such as collective statistical hadronization. This mechanism leads to an increase in the muon production in hadronic cosmic ray interactions~\cite{baur2023core,Manshanden:2022hgf,Werner:2021hdk,Anchordoqui:2022fpn}. The core-corona model of heavy-ion collisions~\cite{werner2007core} combines both production mechanisms: the large-density region of interaction - \textit{core} - hadronizes statistically, while the low-density region - \textit{corona} - is modeled through string fragmentation. 

The muon deficit in air shower simulations is related to the hadronic interaction model's representation of the energy distribution between electromagnetic particles and hadrons, expressed as $ R=E_\mathrm{em} / E_\mathrm{had} $. The neutral to charged pion production ratio $\pi^0/\pi^{\pm}$ strongly impacts $R$, as muons mainly stem from charged pion decay, while neutral pions primarily feed the electromagnetic component. Hadronic interaction simulations implement different hadronization mechanisms to describe the particle population and final state energy, generating different $R$ values.

Current hadronic interaction models use the string fragmentation model to describe hadronization processes, which accurately models electron-positron and low-energy proton-proton collisions. However, in heavy ion collisions, where energy densities are much higher, a fluid-like behavior with statistical hadronization is expected. This scenario favors the production of heavier particles, reducing the fraction of $\pi^0$ compared to other particle types, thereby lowering the  $R$ value. 

This fluid-like behavior, referred to as collective effects, has been observed in both heavy-ion collisions and in proton-proton collisions~\cite{adams2005experimental, adcox2005formation, arsene2005quark, back2005phobos, chatrchyan2013observation, dusling2016novel, loizides2016experimental}. For heavy ions, the formation of a quark-gluon-plasma, described by the laws of hydrodynamics and followed by statistical decay, is modeled as a phase of parton matter where confinement is not required~\cite{shuryak1980quantum, stoecker1986high, kolb2004hydrodynamic}. For protons, the energy densities in central collisions may be large enough to create a quark-gluon-plasma, as well~\cite{werner2011ridge, nagle2018small}. Furthermore, microscopic effects in string fragmentation~\cite{bierlich2018collectivity} and quantum chromodynamics interference~\cite{blok2017collectivity} can also produce collective effects.

A decrease in the value of $R$ in hadronic interactions in proton-proton collisions could solve the muon puzzle~\cite{baur2023core}, as it would result from a more abundant kaon and/or (anti)baryon production. However, collider data and hadronic interaction models constrain $R$: string fragmentation leads to $R\approx0.4$, while statistical models suggest $R\approx0.34$~\cite{alice2017enhanced}. As we will show in the present section, the core-corona approach leads to a lower value of $R$.

\subsection{Implementation in the CONEX framework} \label{subsec:CCimplementation}

We implement a simplified version of the core-corona model~\cite{baur2023core} by changing the energy spectra of produced particles (see Sec.~\ref{subsec:conex}): the yield of the particle species $i$, $$N_i = \fcore \, N_i^{\mathrm{core}} + (1-\fcore) \, N_i^{\mathrm{corona}}, $$ has a contribution $N_i^{\mathrm{core}}$ from the statistical hadronization happening in the core and a contribution $N_i^{\mathrm{corona}}$ from the hadronization through string fragmentation in the corona region. We define $\omega_\mathrm{core}$ as energy dependent (see equation \ref{eq:wcore} below).

In this approach, we only modify the hadronization, disregarding particle correlations from collective effects in the core, which we expect to be negligible. Collective effects can only influence the transversal momentum of particles, which is of small importance at high energies. We treat nuclei following the simplified superposition model. This simplification neglects nuclear effects, which would lead to a stronger core effect. Consequently, this simplification only implies a more conservative version of the core-corona approach. Furthermore, since core hadronization is experimentally demonstrated at mid-rapidity but not excluded for large rapidities, we uniformly apply the core-corona effect at all pseudorapidities, except for the leading particle, whose properties should not be modified. With the same argument, we apply the core-corona effect to all types of hadronic projectiles (nucleons, pions, and kaons). These approximations correspond to an upper limit on what to expect from a core formation for a given value of $\fcore$. Finally, the core weight $\fcore$ needs to increase monotonically with multiplicity to reflect collider data~\cite{alice2017enhanced}, starting from zero for low-multiplicity proton-proton scattering and reaching unity for central Pb-Pb collisions. Since in CONEX tables the multiplicity of each hadronic interaction is not known event by event, but the mean multiplicity increases with the energy of the interaction~\cite{baur2023core}, we can assume that the mean core fraction also increases monotonically with the energy:
\begin{equation}\label{eq:wcore}\fcore(E;\EsubText{acc},E_\mathrm{scale},f_\omega  ) = f_\omega \, \frac{\ln(E/E_\mathrm{acc})}{\ln(E_\mathrm{scale}/E_\mathrm{acc})} \Theta(E-E_\mathrm{acc}).
\end{equation}
The Heaviside step function, $\Theta$, ensures no core effects at energies below $E_\mathrm{acc}$, where models are well constrained by accelerator data. We use a low threshold value of $E_\mathrm{acc} = \SI{100}{\GeV}$, since experimental data beyond fixed target measurements do not well constrain particle production in the relevant phase space. The parameter $E_\mathrm{scale}$ is a reference energy scale, where the modification scale $f_\omega$ is equal to the core fraction $\fcore$.

Since the energy spectra of secondary particles define the features of particle interactions, any modification to the energy spectra used in CONEX's cascade equations directly impacts the air shower simulation. We implement the core-corona approach by modifying these spectra, without changing the total amount of particles produced in the interactions. Each hadronic interaction model in CONEX has its own corona-type particle ratios for each energy bin, projectile, and secondary type. Referring to \cite{andronic2017hadron}, we compute the respective core-type particle ratios using measurements of central Pb-Pb collisions, where we expect to have a $100\%$ core effect. This way, we obtain the ratios $\pi^0 / \pi^\pm$, $\mathrm{p} / \pi^\pm$, $K^\pm / \pi^\pm$, p/n and $K^0 / \pi^\pm$ that represent the core contribution. We directly obtain the charged pion, charged kaon, and proton yields from~\cite{andronic2017hadron}, and we add the decay products from the short-lived $\phi$, $\Lambda$, $\Xi$ and $\Omega$ particles~\cite{andronic2017hadron} following the corresponding branching ratios. Finally, we deduce the neutral pion and neutron yields and compute the ratios using isospin invariance, given that the core- and corona-type ratios are independent of the interaction energy.

For each hadronic model, primary type and interaction energy, the total number of secondary particles at mid-rapidity is $N_\mathrm{tot} = N_{\pi^{0}}+N_{\pi^{\pm}}+N_{K^{\pm}}+N_{K^{0}}+N_\mathrm{p}+N_\mathrm{n}$, which needs to remain unchanged in the core-corona implementation. We can rewrite $N_\mathrm{tot} = A N_{\pi^{\pm}}$, where $ A = 1 + R_{\pi^{0}/\pi^{\pm}} + R_{K^{\pm}/\pi^{\pm}} + R_{K^{0}/\pi^{\pm}} + R_{\mathrm{p}/\pi^{\pm}}  + R_{\mathrm{n}/\mathrm{p}}R_{\mathrm{p}/\pi^{\pm}}$, and $R_{\alpha/\beta}=N_{\alpha}/N_{\beta}$ is the ratio for particle species $\alpha$ and $\beta$. Suppose that core-type hadronization produces all particles. In that case, the scale factors for the secondary particle spectra are $f_\alpha = \tilde{N}_\alpha / N_\alpha$, where $N_\alpha$ is the corona-type yield of the secondary particle type $\alpha$ and $\tilde{N}_\alpha$ is the corresponding yield from the core~\cite{andronic2017hadron}. In an intermediate scenario, where $\omega_\mathrm{core}$ is between 0 and 1, the corresponding scale factor is $f_{\alpha,\omega} = 1 + (f_\alpha-1) \; \omega_\mathrm{core}$. This way, we recover $f_{\alpha,\omega} = 1$ if $\omega_\mathrm{core} = 0$ and $f_{\alpha,\omega} = f_\alpha$ if $\omega_\mathrm{core} = 1$. To ensure that we preserve $N_\mathrm{tot}$ with the new particle ratios, we compute the new core-corona yields (denoted by the superscript $^\mathrm{cc}$) in the following order:
\begin{alignat*}{1}
N^\mathrm{cc}_{\pi^{\pm}} &= N_\mathrm{tot} / A^\mathrm{cc}  \\
N^\mathrm{cc}_{\pi^{0}} &= R^\mathrm{cc}_{\pi^{0}/\pi^{\pm}} \, N^\mathrm{cc}_{\pi^{\pm}}  \\
N^\mathrm{cc}_{K^{\pm}}   &= R^\mathrm{cc}_{K^{\pm}/\pi^{\pm}} \, N^\mathrm{cc}_{\pi^{\pm}}  \\
N^\mathrm{cc}_{K^{0}}     &= R^\mathrm{cc}_{K^{0}/\pi^{0}}     \, N^\mathrm{cc}_{\pi^{0}} \\
N^\mathrm{cc}_p           &= R^\mathrm{cc}_{p/\pi^{\pm}}       \, N^\mathrm{cc}_{\pi^{\pm}}  \\
N^\mathrm{cc}_{n}         &= R^\mathrm{cc}_{n/\pi^{0}}         \, N^\mathrm{cc}_{\pi^{0}}. 
\end{alignat*}
For computing $N^\mathrm{cc}_{\pi^{\pm}}$, we need $N_\mathrm{tot}$ and $A^\mathrm{cc} = 1 + R^\mathrm{cc}_{\pi^{0}/\pi^{\pm}} + R^\mathrm{cc}_{K^{\pm}/\pi^{\pm}} + R^\mathrm{cc}_{K^{0}/\pi^{\pm}} + R^\mathrm{cc}_{\mathrm{p}/\pi^{\pm}}  + R^\mathrm{cc}_{\mathrm{n}/\mathrm{p}}R^\mathrm{cc}_{\mathrm{p}/\pi^{\pm}}$. We use the yields of the modified spectra to compute the new ratios $R^\mathrm{cc}_{\alpha/\beta}$. 

\begin{figure*}[t]
	\centering
 \includegraphics[width=1.0\textwidth]{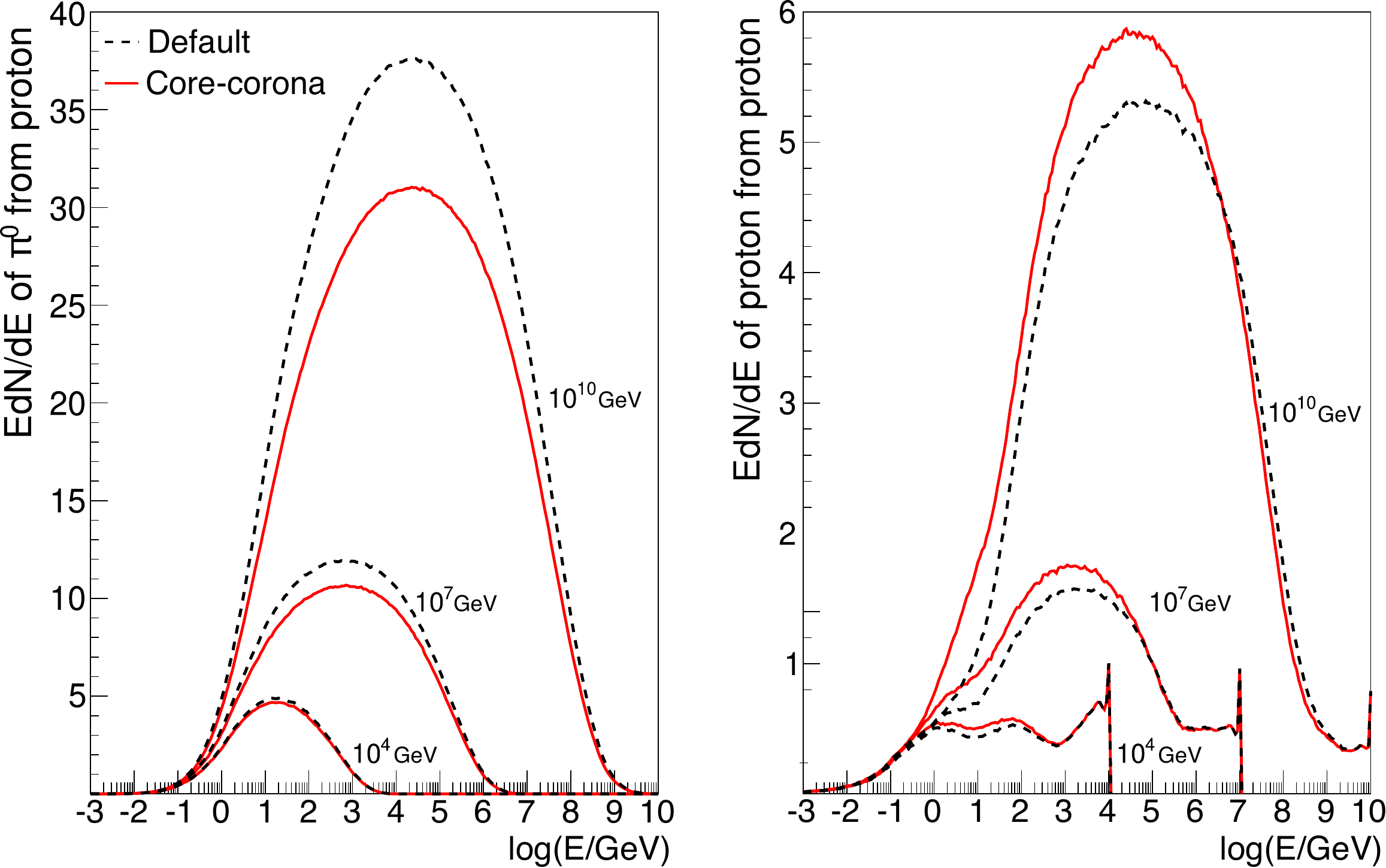}
  \caption{Default (dashed black) and modified (solid red) spectra of secondary $\pi^0$ (left) and protons (right) for proton-air collisions at three different projectile energies and using \QGSJETIId{}. The diffractive peak close to the projectile energy for proton secondaries reflects the leading particle effect.}
  \label{fig:SpectraChanges}
\end{figure*}

Current hadronic interaction models implement different types of hadronization for the central part of the collision and the remnants~\cite{ayala2020core}. String fragmentation describes the central part where most particles are produced at mid-rapidity, corresponding to a broad peak in the center of the energy spectra. The core-corona model applies to this region of interaction. Fig.~\ref{fig:SpectraChanges} shows the neutral pion (left) and proton (right) energy spectra for three different proton-initiated shower energies with the default \QGSJETIId{} hadronic interaction model (dashed black) and with an implementation of the core-corona model in \QGSJETIId{} (solid red). The core-corona implementation leads to a decrease in the neutral pion spectra and an increase in the proton spectra.

\begin{figure}[t] 
	\centering
  \includegraphics[width=1.0\textwidth]{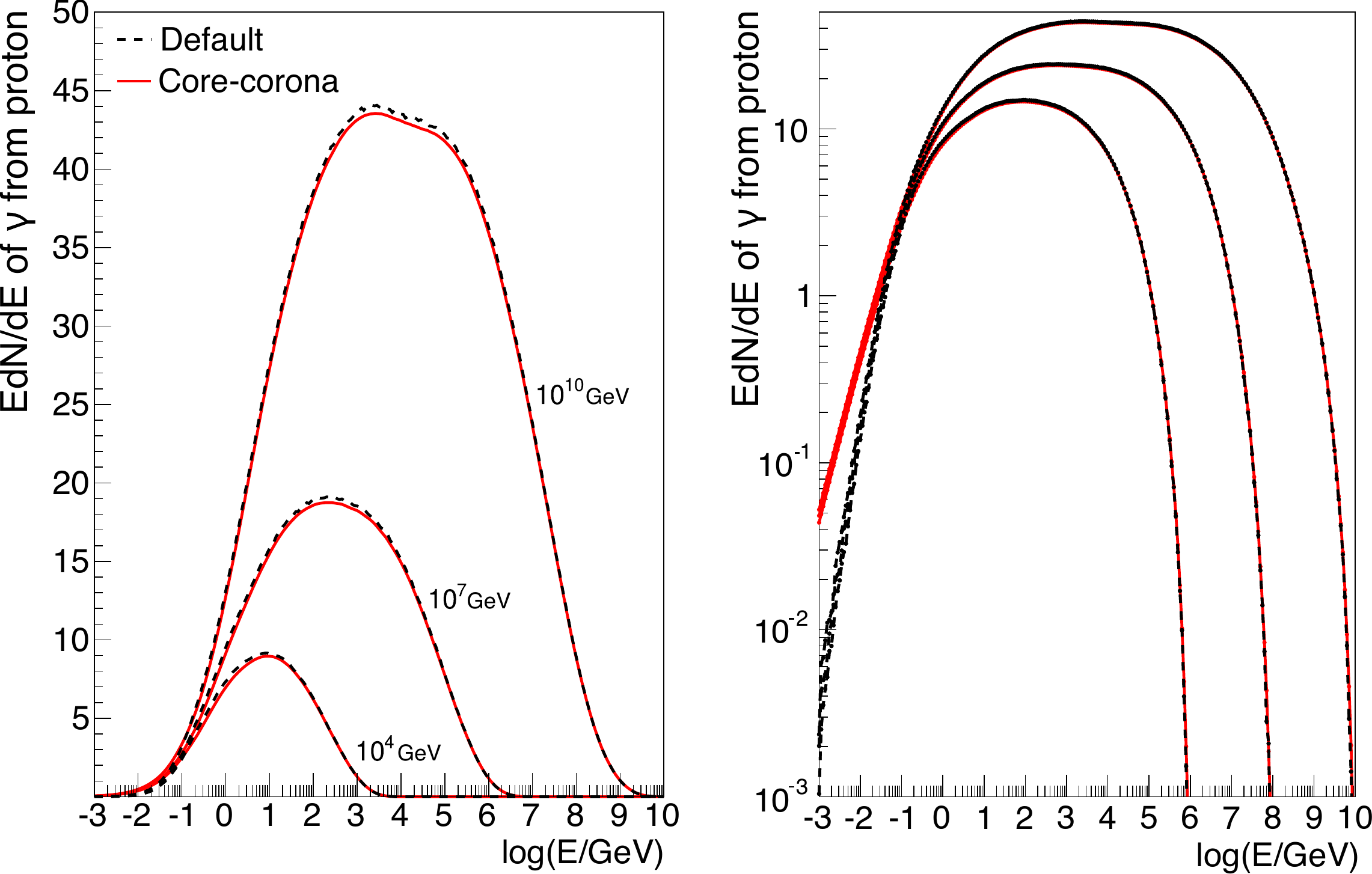}
	\caption{Energy spectra of photons arising from proton-air interactions with different energies and using \QGSJETIId{}, with and without the core-corona effect. Left: linear scale, right: logarithmic scale.}
	\label{gamma-pi0} 
\end{figure}

The remnant hadronization mainly contributes when the secondary and projectile are the same particle type, giving rise to the \emph{leading particle effect} (see Fig.~\ref{fig:SpectraChanges}, right). The corresponding spectrum has a \textit{diffractive peak} at the projectile energy. Since the core-corona model does not affect the remnants, the diffractive peak must remain unaltered in the core-corona implementation. We modify spectra with a leading particle contribution conserving the total energy while leaving the diffractive peak untouched.

CONEX uses photon spectra derived from neutral pions in its electromagnetic cascade equations. Since the core-corona model modifies these neutral pion spectra, we must also compute the new photon spectra. Considering that the decay $\pi^0 \rightarrow \gamma \gamma$ is isotropic in the rest frame, the distribution of produced photons is flat as a function of $\cos \theta^*$:$$\frac{\mathrm{d} N}{\mathrm{d} \cos\theta^*} = \frac{1}{2},$$ where $\theta^*$ is the angle between the $\pi^0$ momentum in the laboratory frame and the photon momentum in the rest frame. We obtain the energy distribution of photons coming from a $\pi^0$ with momentum $p_{\pi}$ from the transformation between the rest and laboratory frames: $$\frac{\mathrm{d} N}{\mathrm{d} E_{\gamma}} = \frac{1}{2} \, \frac{\mathrm{d} \cos \theta^*}{\mathrm{d} E_\gamma} = \frac{2}{p_{\pi}}. $$ We then calculate the number of photons with energy $E_{\gamma}$ by integrating $2/p_{\pi}$ over all energies above $E_{\gamma}$. Finally, we apply a scale factor to the photons to ensure energy conservation after computing the new photon spectra based on the modified neutral pion spectra. We show photon spectra obtained through this procedure, together with the original spectra, in Fig.~\ref{gamma-pi0}. The differences between them at the highest energies produce a small change of $\pm$\SI{2}{\g/\cm^2} in $X_\mathrm{max}$, which can be corrected for in each interaction model, if needed.

\subsection{Air shower simulations}

To assess the impact of the core-corona model on the muon-related observables, we focus on proton-initiated showers with $E_\mathrm{scale} = \SI{e10}{\GeV}$ and $f_\omega=1$. This leads to an $\omega_\mathrm{core}$ value that increases logarithmically from $0$ at $E = \SI{100}{\GeV}$ to a value of 1 at $E = \SI{e10}{\GeV}$ (see Eq.~\ref{eq:wcore}). It is worth noting that the $\omega_\mathrm{core}$ evolution with energy represents a rather strong contribution due to the core. However, it leads to a production of muons in showers with a number compatible with observations~\cite{baur2023core}.

\begin{figure}[t]
	\centering
    \includegraphics[width=1.0\textwidth]{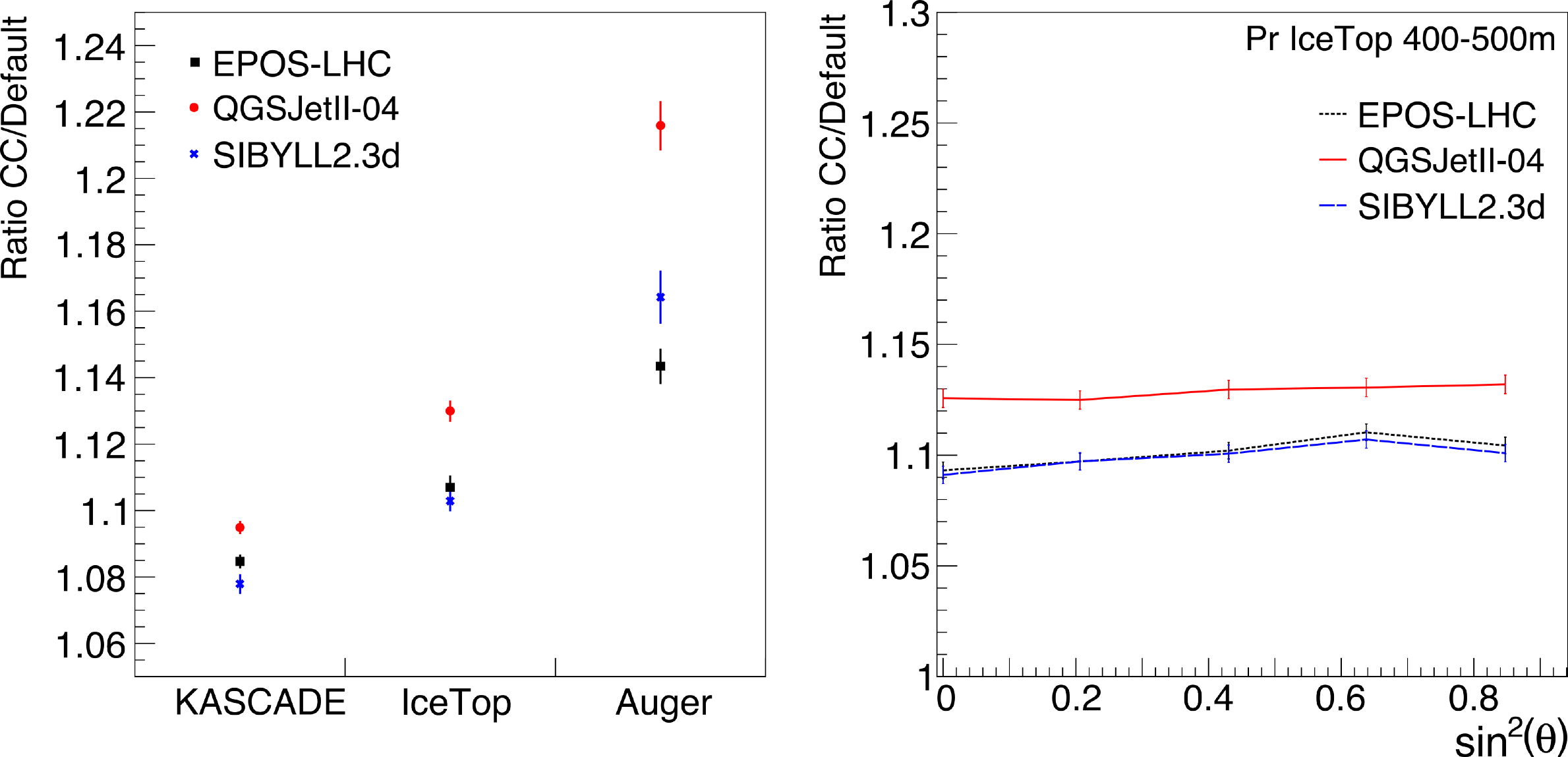}
	\caption{Mean ratios of the total number of muons in proton-initiated showers between the default hadronic interaction model and the corresponding core-corona implementation. We focus on vertical showers at different experiments (left) and differently inclined showers at IceTop (right).}
    \label{fig:CC}
\end{figure} 

Fig.~\ref{fig:CC} (left panel) shows the mean ratio of the muon number for a hadronic interaction model between the core-corona and default versions. We present results for \EPOSLHC{} (black squares), \QGSJETIId{} (red circles), and \SIBYLLd{} (blue crosses) in the KASCADE, IceTop, and Auger scenarios. The core-corona model increases the muon number almost independently of the shower development stage (longitudinal profile), distance to the shower core, and energy of the muon at the ground. The ratio is highest at Auger ($15\%$ to $20\%$) and lowest at KASCADE (about $10\%$), which results from the primary cosmic ray energies considered in each case: the increase in the number of muons in the core-corona model scales with the primary particle energy. Finally, since \QGSJETIId{} produces the lowest number of muons without modification, the impact of the core contribution is here the most significant. The evolution of the ratio with the incidence angle is rather flat in all experiments and hadronic interaction models. We display an example of the IceTop scenario in the right panel of Fig.~\ref{fig:CC}.

This implementation of the core-corona model introduces a global shift in the number of muons, since we only modify the particle ratios but not the spectral shapes. The final number of muons depends on all the generations of the hadronic shower. In this approach, the modifications of the hadronic interactions are significant only at high energy when no 3D development occurs. When the energy is low enough to allow for a lateral spread, the hadronic interactions are not modified anymore. As a consequence, there is less energy going into the electromagnetic shower at high energy, allowing more muon production at low energy, but without a specific signature in the lateral distribution.

The effects of implementing the core-corona model at the Monte Carlo level would be more intricate, as they can modify the transverse momentum of secondary particles. However, we anticipate minimal differences due to the forward boost, which mitigates the influence of these modifications. The primary particle type may have a larger impact, as heavier projectiles tend to produce a stronger core effect. Nevertheless, this would affect only the first interaction, limiting its overall impact. The core formation leads to a maximum increase of $15\%$ to $20\%$ in the number of muons under optimistic conditions. While this is a notable effect, it is insufficient to fully account for the Auger data, unless $X_\mathrm{max}$ is also modified \cite{abdul2024testing}.

\section{Conclusions}\label{sec:conclusion}

In this paper, we presented and validated the implementation of a multidimensional extension to the CONEX simulation framework. This improved framework opens the door for a detailed study of muon-related air shower observables at different cosmic-ray experiments. Additionally, we extended the framework to incorporate the core-corona model, investigating its impact on muon production in air showers to better understand the discrepancies between experimental data and theoretical predictions.

We first addressed air shower simulations, an essential tool for interpreting experimental data, by demonstrating that the CONEX framework significantly reduces computational costs compared to CORSIKA. Additionally, we introduced the CONEX~3D extension, which enables the study of the lateral shower development—particularly the muon component—and showed that it reproduces full Monte Carlo simulations. This progress paves the way for performing more detailed simulations in low computing times.

Then, we applied the extended CONEX framework to study the interplay between various muon-related observables and the characteristics of three cosmic ray experiments sensitive to different shower properties: the Pierre Auger Observatory, the KASCADE Experiment, and the IceTop Array at IceCube. We examined how differences in the muon observables—resulting from the geometry and sensitivities of each experiment—contribute to the model-data discrepancies. For this purpose, we analyzed the longitudinal and lateral distributions of muons, as well as the muon energy spectra. We also compared the cases of proton and iron-initiated showers.

Finally, leveraging the CONEX~3D fast and flexible structure, we presented the implementation of an effective core-corona model. This theory proposes that collective effects in proton-proton collisions significantly impact the hadronic shower content, partially explaining the muon deficit in air showers. We showed that underestimating the contribution of statistical hadronization in hadron-nuclei collisions may indeed lead to a muon deficit in simulations. In our implementation of the core-corona model, the core effect in the hadronic interactions can be changed in an energy-dependent way. We showed that the muon ratio between the default hadronic interaction models and the corresponding core-corona implementations increases with the primary cosmic ray energy, similar to the discrepancy observed between air shower experiments and models. At a primary energy of $E_0 = \SI{e19}{\eV}$, we obtained an increase of $15\%$ to $20\%$ in the muon content, considering a high contribution due to the core. Consequently, to explain the muon deficit observed with the current hadronic interaction models, various models could be considered. However, since the effect of collective hadronization was observed at the LHC, including the core-corona model in hadronic interactions is necessary before considering more exotic scenarios. 

\acknowledgments
The authors would like to thank Nikolai Kalmikov, Vladimir Chernatkin and Klaus Werner for fruitful discussions and their initial work on the electromagnetic cascade equations and Sergey Ostapchenko for his first work on the muon 3D propagation in CONEX.
\printbibliography

@article{Stanev:1994pf,
    author = "Stanev, T. and Vankov, H. P.",
    title = "{Hybrid simulations of electromagnetic cascades}",
    doi = "10.1016/0927-6505(94)90016-7",
    journal = "Astroparticle Physics",
    volume = "2",
    pages = "35--42",
    year = "1994"
}

@article{Manshanden:2022hgf,
    author = {Manshanden, Julien and Sigl, G\"unter and Garzelli, Maria V.},
    title = "{Modeling strangeness enhancements to resolve the muon excess in cosmic ray extensive air shower data}",
    eprint = "2208.04266",
    archivePrefix = "arXiv",
    primaryClass = "hep-ph",
    doi = "10.1088/1475-7516/2023/02/017",
    journal = "JCAP",
    volume = "02",
    pages = "017",
    year = "2023"
}

@article{Werner:2021hdk,
    author = "Werner, K. and Pierog, T. and Guiot, B. and Jahan, J.",
    title = "{Recent Developments in EPOS: Core\textendash{}Corona Effects in Air Showers}",
    doi = "10.1134/S106377882113041X",
    journal = "Physics of Atomic Nuclei",
    volume = "84",
    number = "6",
    pages = "1026--1029",
    year = "2021"
}

@article{Anchordoqui:2022fpn,
    author = "Anchordoqui, Luis A. and Canal, Carlos Garcia and Kling, Felix and Sciutto, Sergio J. and Soriano, Jorge F.",
    title = "{An explanation of the muon puzzle of ultrahigh-energy cosmic rays and the role of the Forward Physics Facility for model improvement}",
    eprint = "2202.03095",
    archivePrefix = "arXiv",
    primaryClass = "hep-ph",
    reportNumber = "DESY-22-021",
    doi = "10.1016/j.jheap.2022.03.004",
    journal = "Journal of High Energy Astrophysics",
    volume = "34",
    pages = "19--32",
    year = "2022"
}

@article{Riehn:2019jet,
    author = "Riehn, Felix and Engel, Ralph and Fedynitch, Anatoli and Gaisser, Thomas K. and Stanev, Todor",
    title = "{Hadronic interaction model Sibyll 2.3d and extensive air showers}",
    eprint = "1912.03300",
    archivePrefix = "arXiv",
    primaryClass = "hep-ph",
    doi = "10.1103/PhysRevD.102.063002",
    journal = "Physical Review D",
    volume = "102",
    number = "6",
    pages = "063002",
    year = "2020"
}

@article{Chernatkin:2003rm,
    author = "Chernatkin, V. and Kalmykov, N. and Werner, K.",
    editor = "Kaidalov, A. B. and Vysotsky, M. I.",
    title = "{Three dimensional electron photon cascade}",
    doi = "10.1080/01422410310001610383",
    journal = "Surveys in High Energy Physics",
    volume = "18",
    pages = "183--186",
    year = "2003"
}

@phdthesis{Chernatkin:2005,
    author = "Chernatkin, V.",
    title = "{Simulation des gerbes dues aux rayons cosmiques a tr\`es hautes \'energies}",
    school = "Universit\'e de Nantes",
    year = "2005"
}

@article{Drescher:2002cr,
    author = "Drescher, Hans-Joachim and Farrar, Glennys R.",
    title = "{Air shower simulations in a hybrid approach using cascade equations}",
    eprint = "astro-ph/0212018",
    archivePrefix = "arXiv",
    doi = "10.1103/PhysRevD.67.116001",
    journal = "Physical Review D",
    volume = "67",
    pages = "116001",
    year = "2003"
}

@inproceedings{Pierog:2011kzx,
    author = "Pierog, Tanguy and Engel, Ralph and Heck, Dieter and Ulrich, Ralf",
    title = "{3D Hybrid Air Shower Simulation in CORSIKA}",
    booktitle = "{32nd ICRC}",
    doi = "10.7529/ICRC2011/V02/1170",
    volume = "2",
    pages = "222",
    year = "2011"
}

@article{albrecht2022muon,
  title={The Muon Puzzle in cosmic-ray induced air showers and its connection to the Large Hadron Collider},
  author={Albrecht, Johannes and Cazon, Lorenzo and Dembinski, Hans and Fedynitch, Anatoli and Kampert, Karl-Heinz and Pierog, Tanguy and Rhode, Wolfgang and Soldin, Dennis and Spaan, Bernhard and Ulrich, Ralf and others},
  journal={Astrophysics and Space Science},
  volume={367},
  number={3},
  pages={27},
  year={2022},
  publisher={Springer}
}

@article{Cazon:2020zhx,
    author = "Cazon, Lorenzo",
    collaboration = "EAS-MSU, IceCube, KASCADE Grande, NEVOD-DECOR, Pierre Auger, SUGAR, Telescope Array, Yakutsk EAS Array",
    title = "{(for EAS-MSU, IceCube, KASCADE Grande, NEVOD-DECOR, Pierre Auger, SUGAR, Telescope Array and Yakutsk EAS Array collaborations) Working Group Report on the Combined Analysis of Muon Density Measurements from Eight Air Shower Experiments}",
    eprint = "2001.07508",
    archivePrefix = "arXiv",
    primaryClass = "astro-ph.HE",
    doi = "10.22323/1.358.0214",
    journal = "PoS",
    volume = "ICRC2019",
    pages = "214",
    year = "2020"
}

@article{matthews2005heitler,
  title={A Heitler model of extensive air showers},
  author={Matthews, James},
  journal={Astroparticle Physics},
  volume={22},
  number={5-6},
  pages={387--397},
  year={2005},
  publisher={Elsevier}
}

@article{kampert2012measurements,
  title={Measurements of the cosmic ray composition with air shower experiments},
  author={Kampert, Karl-Heinz and Unger, Michael},
  journal={Astroparticle Physics},
  volume={35},
  number={10},
  pages={660--678},
  year={2012},
  publisher={Elsevier}
}

@book{heitler1984quantum,
  title={The quantum theory of radiation},
  author={Heitler, Walter},
  year={1984},
  publisher={Courier Corporation}
}

@inproceedings{hillas1981two,
  title={Two interesting techniques for Monte-Carlo simulation of very high energy hadron cascades},
  author={Hillas, AM},
  booktitle={In: ICRC, 17th, Paris, France, July 13-25, 1981, Conference Papers. Volume 8.},
  volume={8},
  pages={193--196},
  year={1981}
}

@inproceedings{muller2019direct,
  title={Direct measurement of the muon density in air showers with the Pierre Auger Observatory},
  author={M{\"u}ller, Sarah},
  booktitle={EPJ Web of Conferences},
  volume={210},
  pages={02013},
  year={2019},
  organization={EDP Sciences}
}

@article{meurer2006muon,
  title={Muon production in extensive air showers and its relation to hadronic interactions},
  author={Meurer, Christine and Bl{\"u}mer, J and Engel, R and Haungs, A and Roth, M},
  journal={Czechoslovak Journal of Physics},
  volume={56},
  pages={A211--A219},
  year={2006},
  publisher={Springer}
}

@article{pierre2020pierre,
  author={Aab, Alexander and others},
  title={The Pierre Auger Observatory and its upgrade},
  collaboration={Pierre Auger},
  journal={Science Reviews-from the end of the world},
  volume={1},
  number={4},
  pages={8--33},
  year={2020}
}

@article{aab2016pierre,
  title={The Pierre Auger Observatory upgrade-preliminary design report},
  author={Aab, Alexander and Abreu, P and Aglietta, MARCO and Ahn, EJ and Samarai, I Al and Albuquerque, IFM and Allekotte, I and Allison, P and Almela, A and Castillo, J Alvarez and others},
  journal={arXiv preprint arXiv:1604.03637},
  year={2016}
}

@article{abbasi2013icetop,
  title={IceTop: The surface component of IceCube},
  author={Abbasi, Rasha and Abdou, Yasser and Ackermann, M and Adams, J and Aguilar, JA and Ahlers, M and Altmann, D and Andeen, K and Auffenberg, J and Bai, X and others},
  journal={Nuclear Instruments and Methods in Physics Research Section A: Accelerators, Spectrometers, Detectors and Associated Equipment},
  volume={700},
  pages={188--220},
  year={2013},
  publisher={Elsevier}
}

@inproceedings{dedenko1965new,
  title={A new method of solving the nuclear cascade equation.},
  author={Dedenko, LG},
  booktitle={ICRC},
  volume={2},
  pages={662},
  year={1965}
}

@article{adams2005experimental,
  title={Experimental and theoretical challenges in the search for the quark gluon plasma: The STAR Collaboration's critical assessment of the evidence from RHIC collisions},
  author={Adams, John and Aggarwal, MM and Ahammed, Z and Amonett, J and Anderson, BD and Arkhipkin, D and Averichev, GS and Badyal, SK and Bai, Y and Balewski, J and others},
  journal={Nuclear Physics A},
  volume={757},
  number={1-2},
  pages={102--183},
  year={2005},
  publisher={Elsevier}
}

@article{adcox2005formation,
  title={Formation of dense partonic matter in relativistic nucleus--nucleus collisions at RHIC: experimental evaluation by the PHENIX collaboration},
  author={Adcox, K and Adler, SS and Afanasiev, S and Aidala, C and Ajitanand, NN and Akiba, Y and Al-Jamel, A and Alexander, J and Amirikas, R and Aoki, K and others},
  journal={Nuclear Physics A},
  volume={757},
  number={1-2},
  pages={184--283},
  year={2005},
  publisher={Elsevier}
}

@article{back2005phobos,
  title={The PHOBOS perspective on discoveries at RHIC},
  author={Back, BB and Baker, MD and Ballintijn, M and Barton, DS and Becker, B and Betts, RR and Bickley, AA and Bindel, R and Budzanowski, A and Busza, W and others},
  journal={Nuclear Physics A},
  volume={757},
  number={1-2},
  pages={28--101},
  year={2005},
  publisher={Elsevier}
}

@article{arsene2005quark,
  title={Quark gluon plasma and color glass condensate at RHIC. The perspective from the BRAHMS experiment},
  author={Arsene, I and Bearden, IG and Beavis, D and Besliu, C and Budick, B and B{\o}ggild, H and Chasman, C and Christensen, CH and Christiansen, P and Cibor, J and others},
  journal={Nuclear Physics A},
  volume={757},
  number={1-2},
  pages={1--27},
  year={2005},
  publisher={Elsevier}
}

@article{chatrchyan2013observation,
  title={Observation of long-range, near-side angular correlations in proton--lead collisions at the LHC},
  author={Chatrchyan, Serguei and Khachatryan, V and Sirunyan, Albert M and Tumasyan, A and Adam, W and Aguilo, E and Bergauer, T and Dragicevic, M and Er{\"o}, J and Fabjan, C and others},
  journal={Physics Letters B},
  volume={718},
  number={3},
  pages={795--814},
  year={2013},
  publisher={Elsevier}
}

@article{dusling2016novel,
  title={Novel collective phenomena in high-energy proton--proton and proton--nucleus collisions},
  author={Dusling, Kevin and Li, Wei and Schenke, Bj{\"o}rn},
  journal={International Journal of Modern Physics E},
  volume={25},
  number={01},
  pages={1630002},
  year={2016},
  publisher={World Scientific}
}

@article{loizides2016experimental,
  title={Experimental overview on small collision systems at the LHC},
  author={Loizides, Constantin},
  journal={Nuclear Physics A},
  volume={956},
  pages={200--207},
  year={2016},
  publisher={Elsevier}
}

@article{shuryak1980quantum,
  title={Quantum chromodynamics and the theory of superdense matter},
  author={Shuryak, Edward V},
  journal={Physics Reports},
  volume={61},
  number={2},
  pages={71--158},
  year={1980},
  publisher={Elsevier}
}

@article{stoecker1986high,
  title={High energy heavy ion collisions—probing the equation of state of highly excited hardronic matter},
  author={Stoecker, Horst and Greiner, Walter},
  journal={Physics Reports},
  volume={137},
  number={5-6},
  pages={277--392},
  year={1986},
  publisher={Elsevier}
}

@article{kolb2004hydrodynamic,
  title={Hydrodynamic description of ultrarelativistic heavy-ion collisions},
  author={Kolb, Peter F and Heinz, Ulrich},
  journal={Quark--Gluon Plasma 3},
  pages={634--714},
  year={2004},
  publisher={World Scientific}
}

@article{werner2011ridge,
  title={The “Ridge” in Proton-Proton Scattering at 7~TeV},
  author={Werner, Klaus and Karpenko, Iu and Pierog, T},
  journal={Physical Review Letters},
  volume={106},
  number={12},
  pages={122004},
  year={2011},
  publisher={APS}
}

@article{bierlich2018collectivity,
  title={Collectivity without plasma in hadronic collisions},
  author={Bierlich, Christian and Gustafson, G{\"o}sta and L{\"o}nnblad, Leif},
  journal={Physics Letters B},
  volume={779},
  pages={58--63},
  year={2018},
  publisher={Elsevier}
}

@article{blok2017collectivity,
  title={Collectivity from interference},
  author={Blok, Boris and J{\"a}kel, Christian D and Strikman, Mark and Wiedemann, Urs Achim},
  journal={Journal of High Energy Physics},
  volume={2017},
  number={12},
  pages={1--50},
  year={2017},
  publisher={Springer}
}

@article{nagle2018small,
  title={Small system collectivity in relativistic hadronic and nuclear collisions},
  author={Nagle, James L and Zajc, William A},
  journal={Annual Review of Nuclear and Particle Science},
  volume={68},
  pages={211--235},
  year={2018},
  publisher={Annual Reviews}
}

@article{alice2017enhanced,
  title={Enhanced production of multi-strange hadrons in high-multiplicity proton--proton collisions},
  journal={Nature Physics},
  volume={13},
  number={6},
  pages={535--539},
  year={2017},
  publisher={Nature Publishing Group UK London}
}

@article{werner2007core,
  title={Core-corona separation in ultrarelativistic heavy ion collisions},
  author={Werner, Klaus},
  journal={Physical Review Letters},
  volume={98},
  number={15},
  pages={152301},
  year={2007},
  publisher={APS}
}

@article{engel2011extensive,
  title={Extensive air showers and hadronic interactions at high energy},
  author={Engel, Ralph and Heck, Dieter and Pierog, Tanguy},
  journal={Annual review of nuclear and particle science},
  volume={61},
  pages={467--489},
  year={2011},
  publisher={Annual Reviews}
}

@article{antoni2003cosmic,
  title={The cosmic-ray experiment KASCADE},
  author={Antoni, T and Apel, WD and Badea, F and Bekk, K and Bercuci, A and Bl{\"u}mer, H and Bozdog, H and Brancus, IM and B{\"u}ttner, C and Chilingarian, A and others},
  journal={Nuclear Instruments and Methods in Physics Research Section A: accelerators, spectrometers, detectors and associated equipment},
  volume={513},
  number={3},
  pages={490--510},
  year={2003},
  publisher={Elsevier}
}

@article{ranchon2005response,
  title={Response of a Pierre Auger Observatory surface detector to MeV electrons and GeV muons},
  author={Ranchon, S and Urban, M},
  journal={Nuclear Instruments and Methods in Physics Research Section A: Accelerators, Spectrometers, Detectors and Associated Equipment},
  volume={538},
  number={1-3},
  pages={483--495},
  year={2005},
  publisher={Elsevier}
}

@article{apel2010kascade,
  title={The KASCADE-grande experiment},
  author={Apel, WD and Arteaga, JC and Badea, AF and Bekk, K and Bertaina, M and Bl{\"u}mer, J and Bozdog, H and Brancus, IM and Buchholz, P and Cantoni, E and others},
  journal={Nuclear Instruments and Methods in Physics Research Section A: accelerators, spectrometers, detectors and associated equipment},
  volume={620},
  number={2-3},
  pages={202--216},
  year={2010},
  publisher={Elsevier}
}

@article{ayala2020core,
  title={Core meets corona: A two-component source to explain $\Lambda$ and $\Lambda^-$ global polarization in semi-central heavy-ion collisions},
  author={Ayala, Alejandro and Torres, Marco Alberto Ayala and Cuautle, Eleazar and Dominguez, Isabel and Sanchez, Marcos Aurelio Fontaine and Maldonado, Ivonne and Moreno-Barbosa, Eduardo and Nieto-Mar{\'\i}n, PA and Rodriguez-Cahuantzi, Mario and Salinas, Jordi and others},
  journal={Physics Letters B},
  volume={810},
  pages={135818},
  year={2020},
  publisher={Elsevier}
}

@article{baur2023core,
  title={Core-corona effect in hadron collisions and muon production in air showers},
  author={Baur, Sebastian and Dembinski, Hans and Perlin, Matias and Pierog, Tanguy and Ulrich, Ralf and Werner, Klaus},
  journal={Physical Review D},
  volume={107},
  number={9},
  pages={094031},
  year={2023},
  publisher={APS}
}

@article{aab2021design,
  title={Design, upgrade and characterization of the silicon photomultiplier front-end for the AMIGA detector at the Pierre Auger Observatory},
  author={Aab, Alexander and Abreu, Pedro and Aglietta, Marco and Albury, Justin M and Allekotte, Ingomar and Almela, Alejandro and Alvarez-Mu{\~n}iz, Jaime and Batista, R Alves and Anastasi, Gioacchino Alex and Anchordoqui, Luis and others},
  journal={Journal of Instrumentation},
  volume={16},
  number={01},
  pages={P01026},
  year={2021},
  publisher={IOP Publishing}
}

@inproceedings{andronic2017hadron,
  title={Hadron yields, the chemical freeze-out and the QCD phase diagram},
  author={Andronic, A and Braun-Munzinger, P and Redlich, K and Stachel, J},
  booktitle={Journal of Physics: Conference Series},
  volume={779},
  number={1},
  pages={012012},
  year={2017},
  organization={IOP Publishing}
}

@inproceedings{hillas1965calculations,
  title={Calculations on the propagation of mesons in extensive air showers},
  author={Hillas, AM},
  booktitle={ICRC},
  volume={2},
  pages={758},
  year={1965}
}

@article{bossard2001cosmic,
  title={Cosmic ray air shower characteristics in the framework of the parton-based Gribov-Regge model NEXUS},
  author={Bossard, G and Drescher, HJ and Kalmykov, NN and Ostapchenko, S and Pavlov, AI and Pierog, T and Vishnevskaya, EA and Werner, K},
  journal={Physical Review D},
  volume={63},
  number={5},
  pages={054030},
  year={2001},
  publisher={APS}
}

@article{bergmann2007one,
  title={One-dimensional hybrid approach to extensive air shower simulation},
  author={Bergmann, Till and Engel, Ralph and Heck, Dieter and Kalmykov, NN and Ostapchenko, Sergey and Pierog, Tanguy and Thouw, T and Werner, Klaus},
  journal={Astroparticle Physics},
  volume={26},
  number={6},
  pages={420--432},
  year={2007},
  publisher={Elsevier}
}

@article{heck1998corsika,
  title={CORSIKA: A Monte Carlo code to simulate extensive air showers},
  author={Heck, Dieter and Knapp, Johannes and Capdevielle, JN and Schatz, G and Thouw, T and others},
  journal={Report fzka},
  volume={6019},
  number={11},
  year={1998}
}

@article{heckextensive,
  title={Extensive Air Shower Simulation with CORSIKA: A User’s Guide (Version 7.7550 from April 30, 2024)},
  author={Heck, D and Pierog, T}
}

@article{abdul2024testing,
  title={Testing hadronic-model predictions of depth of maximum of air-shower profiles and ground-particle signals using hybrid data of the Pierre Auger Observatory},
  author={Abdul Halim, A and Abreu, P and Aglietta, M and Allekotte, I and Almeida Cheminant, K and Almela, A and Aloisio, R and Alvarez-Mu{\~n}iz, J and Ammerman Yebra, J and Anastasi, GA and others},
  journal={Physical Review D},
  volume={109},
  number={10},
  pages={102001},
  year={2024},
  publisher={APS}
}

@article{Pierog:2013ria,
      author         = "Pierog, T. and Karpenko, Iu. and Katzy, J. M. and Yatsenko, E. and Werner, K.",
      title          = "{EPOS LHC: Test of collective hadronization with data measured at the CERN Large Hadron Collider}",
      journal        = "Physical Review C",
      volume         = "92",
      year           = "2015",
      number         = "3",
      pages          = "034906",
      doi            = "10.1103/PhysRevC.92.034906",
      eprint         = "1306.0121",
      archivePrefix  = "arXiv",
      primaryClass   = "hep-ph",
      reportNumber   = "DESY-13-125",
      SLACcitation   = "%%CITATION = ARXIV:1306.0121;%%"
}

@article{Ostapchenko:2010vb,
    author = "Ostapchenko, Sergey",
    title = "{Monte Carlo treatment of hadronic interactions in enhanced Pomeron scheme: I.\ QGSJET-II model}",
    eprint = "1010.1869",
    archivePrefix = "arXiv",
    primaryClass = "hep-ph",
    doi = "10.1103/PhysRevD.83.014018",
    journal = "Physical Review D",
    volume = "83",
    pages = "014018",
    year = "2011"
}

\end{document}